\documentclass[a4paper,11pt]{article}

\pdfoutput=1 

\usepackage{jheppub} 

\usepackage[T1]{fontenc} 

\usepackage{graphicx}  
\usepackage{dcolumn}   
\usepackage{bm}        
\usepackage{amssymb}   
\usepackage{amsmath}   
\usepackage{listings}
\usepackage{array}
\usepackage{graphicx}
\usepackage{lineno}
\usepackage{dcolumn}
\usepackage{color}
\usepackage{overpic}
\usepackage{multirow}
\usepackage{hyperref}
\usepackage{cleveref}

\newcommand{\LQ}{\text{LQ}}
\newcommand{\GeV}{\:\text{GeV}}
\newcommand{\etmiss}{E_{\text{T}}^{\text{miss}}}
\newcommand{\phimiss}{\phi^{\text{miss}}}

\newcommand{\taulep}{\tau_{\text{lep}}}
\newcommand{\tauhad}{\tau_{\text{had}}}
\newcommand{\pt}{p_{\text{T}}}

\newcommand{\mL}{\mathcal{L}}

\title{Search for Low-Mass Leptoquarks using the $llqq$ final states in Pb-Pb Ultra-Peripheral Collisions}
\author{Li-Gang Xia}
\affiliation{School of Physics, Nanjing University, Jiangsu Province, CN, 210000}
\emailAdd{ligang.xia@cern.ch}

\abstract{
    After a review over past experiments and theoretical requirements from low-energy flavour physics, we argue that the possibility of low-mass leptoquarks (LQ) cannot be fully excluded due to the assumptions made in the measurements. Therefore we propose to search for the pair production of low-mass leptoquarks in Pb-Pb ultra-peripheral collisions with the least model dependence (assuming a small coupling constant $\lesssim0.1$). There are a couple of advantages: 1) high photon flux provides high production rate for low mass LQs; 2) the background contamination is much lower than that in $p$-$p$ collisions. 
    The analysis strategy permits a leptoquark to decay to all possible lepton-plus-quark modes.
    Taking the scalar LQ, $S_3$, with an electric charge $|q|=\frac{4}{3}e$, as example, the mass point of 100~GeV can be excluded at the 95~\% confidence level using a dataset of 4~pb$^{-1}$ Pb-Pb ultra-peripheral collisions at $\sqrt{s}=5.02$~TeV and the performance of the ATLAS detector in Run~2. The proposed method also applies to searching for high-mass LQs in $p$-$p$ collisions as long as the LQ pair production mechanism dominates.
}
\arxivnumber{2012.15618}
\begin{document}
\maketitle 
\flushbottom

\section{Introduction}
Leptoquarks (LQ) are bosons to connect the quark and lepton sectors. They carry fractional electric charge and color. LQs are predicted in many extensions of the Standard Model (SM), such as grand unified theories~\cite{PatiSalam1,PatiSalam2,GeorgiGlashow,Unify1}, technicolor models~\cite{Buchmuller,DStechnicolor,Dtechnicolor,FStechnicolor,LRtechnicolor}, composite LQ models~\cite{lightLQ, CompositeLQ1, CompositeLQ2} and other models~\cite{ELdynabreaking,Ellis}.
They are able to explain the evidences of the lepton universality violation (LUV) measured in the $B$-meson decays~\cite{Hiller,BLQ1,BLQ2,BLQ3,BLQ4,BLQ5,BLQ6,BLQ7,BLQ8}. Most recently, the LHCb experiment reported another evidence for LUV in the decays $B^+\to K^+l^+l^-$ with a significance of 3.1~$\sigma$~\cite{LHCbnew}. Introducing one or more LQs seems a natural explanation~\cite{LHCbwater1,LHCbwater2,LHCbwater3,LHCbwater4,LHCbwater5,LHCbwater6,LHCbwater7}.

Various LQs have been searched for in the LEP, Tevatron, LHC and other experiments. 
Ref.~\cite{LQreview} is a good review on the LQ physics and experimental searches up to 2016. 
Table~\ref{tab:summary_searches} summarizes a list of experimental searches from different colliders. 
From the table, it seems that LQs are excluded up to 1~TeV or more. However, all searching results are interpreted with assumptions. One common assumption is that a LQ couples only to the same generation of fermions (this assumption is denoted by ``SG'' in Table~\ref{tab:summary_searches}). For example, a so-called ``third-generation'' LQ with a charge $+\frac{2}{3}e$ only couples to $b\tau^+$ or $t\bar{\nu}_\tau$. 
This assumption is not appropriate because of more and more strong evidences for LUV. Taking the decay $B^+\to K^+l^+l^-$ as in the LHCb measurement~\cite{LHCbnew} as example, a LQ must be able to couple to $\mu/e$ and $b/s$ quark if LQ is the right solution.  
A much looser assumption is that a LQ couples to a single lepton plus all possible quarks (this assumption is denoted by ``SL'' in Table~\ref{tab:summary_searches}). This assumption was used in the OPAL experiment~\cite{OPAL2,OPAL3}. Inspired by their strategy, a model-independent method is proposed in Sec.~\ref{sec:method}. For the measurements using single LQ production processes, we have to interpret the results as a function of the LQ mass and the coupling constant between a LQ and the
fermions (denoted by $\lambda$
or $\lambda_{lq}$ for the lepton $l$ and the quark $q$). One common way is to set constraints on the LQ mass assuming a
coupling strength which is equal to that in the electromagnetic interaction, namely, $\lambda=\sqrt{4\pi\alpha_{em}}$ in Table~\ref{tab:summary_searches}, where $\alpha_{em}\approx\frac{1}{137}$ is the fine structure constant. 
\begin{table}[htbp]
    \centering
    \caption{\label{tab:summary_searches}
        Overview of the experimental searches for the leptoquarks. ``SG'' represents the assumption that a LQ couples to the same generation of fermions. ``SL'' represents the assumption that a LQ couples to single lepton and all possible quarks. It should be noted that the presented exclusion region may be just part of the experiment results. 
    }
        \begin{tabular}{l | l | l}
    \hline
            Experiment & Exclusion region (condition) & Assumption \\
            \hline
            AMY\cite{AMY}& 4.3-26.3~GeV ($|q|=\frac{2}{3}e, B(\LQ \to eq)=1$) & SG, LQ$\to e q/\mu q/\nu q$ ($q=d,s$) \\
            CELLO\cite{CELLO} & 7-20.5~GeV & LQ$\to \mu q/\nu q$ \\
            ALEPH\cite{ALEPH} & 4-44~GeV ($|q|=\frac{1}{3}e$) & SG, LQ$\to eu/\nu_ed$\\
            & 6-44~GeV ($|q|=\frac{1}{3}e$) & SG, LQ$\to \mu c/\nu_\mu s$\\
            DELPHI\cite{DELPHI} & 134~GeV ($B(\LQ\to eq)=1$,$\lambda=\sqrt{4\pi\alpha_{em}}$) & SG, LQ$\to eq/\nu_eq$ ($q=u,d$) \\
            OPAL\cite{OPAL1} & $\lambda<1$ for 200 GeV, ($|q|=\frac{4}{3}$, B(\LQ$\to eq$)=1) & LQ$\to eq/\nu_e q$, ($q=u,d,s,c,b$)\\
            OPAL\cite{OPAL2}& 46.3~GeV ($|q|=\frac{4}{3},B(\LQ\to eq/\mu q)=1$) & LQ$\to e q/\mu q$, ($q=u,d,s,c,b$) \\ 
            & 45.5~GeV  ($|q|=\frac{4}{3},B(\LQ\to \tau q)=1$) & SG, LQ$\to \tau q$ , ($q=u,d,s,c,b$)\\ 
            OPAL\cite{OPAL3} & 50-100~GeV ($|q|=\frac{4}{3}, B(\LQ\to eq)=1$) & SL, LQ $\to eq$, ($q=u,d,s,c,b$)\\
            & 50-101~GeV ($|q|=\frac{4}{3}, B(\LQ\to \mu q)=1$) & SL, LQ $\to \mu q$, ($q=u,d,s,c,b$)\\
            & 50-99~GeV ($|q|=\frac{4}{3}, B(\LQ\to \tau q)=1$) & SL, LQ $\to \tau q$, ($q=u,d,s,c,b$)\\
            H1\cite{H11} & 257~GeV ($B(\LQ\to eq)=1$,$\lambda=\sqrt{4\pi\alpha_{em}}$) &  LQ$\to eq/\nu_eq$ ($q=u,d$) \\
            H1\cite{H12}& 712~GeV ($\lambda=\sqrt{4\pi\alpha_{em}}$) & LQ$\to eq/\mu q$, ($q=u,d,s,c,b$) \\
            H1\cite{H13}& 800~GeV ($B(\LQ\to eq)=1,\lambda=\sqrt{4\pi\alpha_{em}}$) &  LQ$\to eq/\nu_e q$\\
            CDF\cite{CDF0} & 99~GeV (B(\LQ$\to \tau b$)=1) & SG, LQ$\to \tau b$ \\
            CDF\cite{CDF1} & 236~GeV (B(\LQ$\to eq$)=1) & SG, LQ$\to eq/\nu_e q$ ($q=u,d$) \\
            CDF\cite{CDF2} & 224~GeV (B(\LQ$\to \mu q$)=1) & SG, LQ$\to \mu q/\nu_\mu q$ ($q=c,s$) \\
            CDF\cite{CDF3} & 317~GeV (B(Vector LQ$\to \tau b$)=1) & SG, Vector LQ$ \to \tau b$\\
            D0\cite{D03} & 229~GeV (B(\LQ$\to \nu_\tau b$)=1) & SG, LQ$ \to \nu_\tau b$\\
            D0\cite{D05} & 210~GeV (B(\LQ$\to \tau b$)=1) & SG, LQ$ \to \tau b$\\
            D0\cite{D02} & 316~GeV (B(\LQ$\to\mu q$)=1) & SG, LQ$\to\mu q/\nu_\mu q$ ($q=c,s$) \\
            D0\cite{D01}& 295~GeV (B(\LQ$\to eq$)=1) & SG, LQ$\to eq/\nu_e q$ ($q=u,d$) \\
            D0\cite{D04} & 247~GeV (B(\LQ$\to \nu_\tau b$)=1) & SG, LQ$ \to \nu_\tau b$\\
            ATLAS\cite{ATLAS1} & 200-534~GeV (B(\LQ$\to \tau b$)=1) & SG, LQ$ \to \tau b$\\
            ATLAS\cite{ATLAS2} & 200-800~GeV  & SG, LQ$ \to \nu_\tau t/\tau b/\nu_\tau b/\tau t$\\
            ATLAS\cite{ATLAS3} & 0.4-1.8~TeV (B(\LQ$\to eq$)=1) & SL, LQ$ \to e q$ ($q=u,d,s,c,b$) \\
            & 0.4-1.7~TeV (B(\LQ$\to \mu q$)=1) & SL, LQ$ \to e q$ ($q=u,d,s,c,b$)\\
            ATLAS\cite{ATLAS4} & 0.9-1.48~TeV (B(\LQ$\to et$)=1) & SL, LQ $\to et$\\
            & 0.9-1.47~TeV (B(\LQ$\to \mu t$)=1) & SL, LQ $\to \mu t$\\
            ATLAS\cite{ATLAS5} & 0.5-1.43~TeV ($|q|=\frac{1}{3}$,B(\LQ$\to \tau t$)=1) & SG, LQ $\to \tau t/\nu_\tau b$\\
            CMS\cite{CMS1}& 200-740~GeV (B(\LQ$\to \tau b$)=1) & SG, LQ$ \to \tau b/\nu_\tau t$\\
            CMS\cite{CMS2}& 200-1730~GeV (B(\LQ$\to e^- u$)=1, $\lambda=1$) & SG, LQ$ \to e^- u$\\
               & 300-660~GeV (B(\LQ$\to \mu^- c$)=1, $\lambda=1$) & SG, LQ$ \to \mu^- c$\\
            CMS\cite{CMS3}& 200-850~GeV (B(\LQ$\to \tau b$)=1) & SG, LQ$ \to \tau b$\\
            CMS\cite{CMS4} & 200-740~GeV (B(\LQ$\to \tau b$)=1, $\lambda=1$) & SG, LQ$ \to \tau^+b$\\
            CMS\cite{CMS5} & 900~GeV ($|q|=\frac{1}{3},B(\LQ\to t\mu/t\tau/b\nu)=1$) & LQ$\to t\mu/t\tau/b\nu)=1$\\ 
            CMS\cite{CMS6}& 250-1020~GeV (B(\LQ$\to \tau b$)=1) & SG, LQ$ \to \tau b$\\
            CMS\cite{CMS7} & 950~GeV (B(\LQ$\to \tau^- t$)=B(\LQ$\to \nu_\tau b$)=0.5) & SG, LQ$ \to \tau^- t/\nu_\tau b$\\
        \hline
        \end{tabular}
\end{table}

A combined fit to all measurements is desirable to avoid the assumptions, but not available for the moment. In Sec.~\ref{sec:rev}, we will show that 80~GeV is a robust lower limit for the LQ with an electric charge $|q|=\frac{4}{3}e$ based on the OPAL measurements~\cite{OPAL2,OPAL3}. Considering a LQ with an electric charge $|q|=\frac{4}{3}e$, it has 9 decay modes ($\LQ\to l+q$ with $l=e,\mu,\tau$ and $q=d,s,b$).  Except the OPAL experiment, there is no search for the final state with a tau lepton plus light quarks. Therefore, it seems that we cannot deny the possibility of a low-mass LQ based on previous searches
without a global fit. 
On the other hand, we will show that low-energy flavour physics~\cite{theory_bound1,theory_bound2,theory_bound3,theory_bound4} can impose strong constraint on combinations of  $\lambda_{lq}$ and $m_{\LQ}$ in Sec.~\ref{sec:rev}. 
We can only set upper limits on $\lambda_{lq}$ for a fixed LQ mass or vice versa. 
Given these difficulties, model-independent methods to search for LQs are highly necessary. 

For the low mass LQs (less than 200~GeV), we find that the nuclear ultra-peripheral collision (UPC) is an ideal place to perform such searches. In UPCs, relativistic nuclei collide at impact parameters ($b$) that are so large that there are no hadronic interactions. Thus it provides a unique opportunity to study photon-photon interactions. Ref.~\cite{UPCreview1,UPCreview2} are two good reviews about the physics of UPCs. 
Given high photon flux at low energies (proportional to the 4-th power of the nucleus charge) and much cleaner collision environment, we propose to search for the LQ pair production in Pb-Pb UPC events. 
Most of previous searches use the property that LQs carry color charge. This proposal uses the property that LQs have electric charge, and is much less model-dependent compared to the existing searching strategies.
Taking the scalar LQ with an electric charge $|q|=\frac{4}{3}e$, denoted as $S_3$ as in the review~\cite{LQreview}, as example, a feasibility study using all possible $llqq$ ($l=e,\mu,\tau$, $q=d,s,b$ and charge conjugation is always assumed) final states is conducted based on the performance of the ATLAS detector in Run~2 and Pb-Pb UPCs at the center-of-mass (c.m.) energy $\sqrt{s}=5.02$~TeV. 

In Sec.~\ref{sec:rev}, the constraints on the LQ mass and coupling constant from the experiments and the low-energy flavour physics are reviewed. 
In Sec.~\ref{sec:method}, a much less model-dependent method is proposed to search for LQ pair production, which process dominates if the coupling strength of the LQ to the fermions is not big ($\lesssim 0.1$ as shown in Fig.~\ref{fig:xs_lam}). 
It is worth emphasizing that the proposed method itself is also applicable to search for high-mass LQs in $p$-$p$ collisions as long as the pair production mechanism dominates. 
In Sec.~\ref{sec:MC}, we describe the cross sections in UPCs and the preparation of MC samples. In Sec.~\ref{sec:selection}, we elaborate the selection criteria and the analysis strategy. In Sec.~\ref{sec:unc}, we will analyze the dominant systematic uncertainties and propose some solutions. In
Sec.~\ref{sec:stats}, we will describe the statistical interpretation scheme and present the sensitivity results. In Sec.~\ref{sec:ex_m100}, we apply the method to the available $p$-$p$ dataset from ATLAS~\cite{ATLAS2} to investigate the possibility of excluding a 100~GeV LQ unambiguously. Sec.~\ref{sec:summary} is a short summary.

\section{\label{sec:rev} The bounds from the experimental searches and low-energy flavour physics}
\subsection{\label{subsec:OPAL}A robust lower mass bound from the OPAL experiments}
For a LQ, both its mass and coupling constant are unknown. The partial width of $\LQ \to l+q$ is 
\begin{equation}
    \Gamma_{lq} = \frac{\lambda_{lq}^2m}{C\pi} \:,
\end{equation}
where $C=16$ for a scalar LQ and $24$ for a vector LQ. The LQ would decay outside a detector for a mass of $100$~GeV and a coupling constant of less than $10^{-8}$. Since a LQ carries both electric and color charge, a long-lived one would interact with the detector matter and may appear like a $b$-tagged jet. There is no dedicated search for such objects. If the lifetime is not too long, some direct searches are summarized in Table~\ref{tab:summary_searches}. Among these results, two of the OPAL
measurements~\cite{OPAL2,OPAL3} seem to use the weakest assumption, which forbids a LQ to couple to different generations of leptons, but allow to couple to all possible quarks. In addition, it used the pair production in $e^+e^-$ collisions. The pair production dominates if $\lambda \lesssim 10^{-2}$ and its cross section does not depend upon $\lambda$.
Collecting the branching fractions, we can naively interpret their results in a different way (taking the LQ with the electric charge $|q|=\frac{4}{3}e$ and the measurement~\cite{OPAL3} as example).
\begin{eqnarray}
    \sigma(m) B_{ex}^2 &<& \sigma(100\GeV) \:,\\
    \sigma(m) B_{\mu x}^2 &<& \sigma(101\GeV) \:,\\
    \sigma(m) B_{\tau x}^2 &<& \sigma(99\GeV) \:,
\end{eqnarray}
where $\sigma(m)$ is the production cross section of a pair of LQs with a mass $m$ in the process $e^+e^- \to \LQ\overline{\LQ}$; $B_{lx}$ is the branching fraction of $\LQ \to lx$ with $x=d,s,b$ for a fixed lepton $l$. Using $B_{ex}+B_{\mu x}+B_{\tau x}=1$ and the Cauchy-Schwarz inequality, we have
\begin{eqnarray}
    \sigma(m) (B_{ex}+B_{\mu x} + B_{\tau x})^2 \leq && \sigma(m) 3 (B_{ex}^2+B_{\mu x}^2 + B_{\tau x}^2) \\
    < && 3(\sigma(100\GeV) + \sigma(101\GeV)+\sigma(99\GeV)) \\
    < && 9\sigma(99\GeV) \:.
\end{eqnarray}
Using the cross section formula in Ref.~\cite{OPAL3} and the highest collision energy, a conservative bound is obtained.
\begin{equation}
 m > 80\: \text{GeV}
\end{equation}
This interpretation neither considers the shape information of the invariant mass reconstructed from a lepton candidate and a jet, nor considers the fact that there are multiple collision energy points with different luminosities. It does not take into account the contribution from $\LQ \to \tau + q$ with a subsequent leptonical $\tau$ decay in the $eeqq$ or $\mu\mu qq$ signal region, either. A full consideration shall give a better bound (but would not exceed $\sim$100~GeV anyway). Thus we claim this is a robust bound with little model
dependence. 

A recent ATLAS measurement looked at the final states $eq$ and $\mu q$ with $q$ being all possible quarks except the top quark and searched for the LQ mass above 400~GeV~\cite{ATLAS3}. Exclusion regions as a function of the LQ mass and the branching fraction is reported (more details see~\cite{ATLAS3}). Unfortunately, there seems no search for the final state with a tau lepton and light quarks except the OPAL experiment. It is difficult to exclude LQs with higher mass without assumptions.

Other experiments as summarized in Table~\ref{tab:summary_searches} either focused on only specific final state or made very strong assumptions. No further bounds can be obtained without a combined fit.

\subsection{The bounds from low-energy flavour physics}
Low-energy flavour physics like rare meson decays and lepton decays can provide stringent bounds on combinations of the coupling constant and mass~\cite{theory_bound1,theory_bound2,theory_bound3,theory_bound4}. 
In most cases, a LQ propagator would inevitably couple to a lepton and a quark on both ends. The extra amplitude induced by a LQ would be proportional to a factor of $\lambda_{lq}\lambda_{l^\prime q^\prime}/m_{\LQ}^2$ if the involved energy scale is much smaller than the LQ mass. This factor is invariant under the scaling transformation $\lambda \to k\lambda$ and $m_{\LQ} \to k m_{\LQ}$. Here we quote a few bounds from the literature. For $K_L \to e^{\pm}\mu^{\mp}$, Ref.~\cite{theory_bound2} obtained
\begin{equation}
    |\lambda|^2 < 6\times 10^{-7}(\frac{m_{\LQ}}{100\GeV})^2 \: .
\end{equation}
For the $\mu^-\to e^-$ conversion on the Titanium~\cite{mu2e_exp}, Ref.~\cite{theory_bound3} obtained
\begin{equation}
    |\lambda_{\mu q}\lambda_{e q}| < 1.7\times 10^{-7}(\frac{m_{\LQ}}{100\GeV})^2 \: .
\end{equation}
Interestingly, the neutral meson ($K^0,D^0,B^0$) mixing is found to be able to constrain $\lambda^2/m_{\LQ}$. A recent calculation~\cite{theory_bound4} gives roughly (the exact bound is model-dependent)
\begin{equation}
    |\lambda|^2 < 2.7\times 10^{-3}\frac{m_{\LQ}}{100\GeV} \: .
\end{equation}

Due to the scaling invariance above, one cannot provide definite bounds on a LQ's mass due to the mixing with the coupling constant. 
The existing anomalies in $B$-meson decays need significant contribution from new physics. The bounds based on those anomalies are highly model-dependent. We shall not expand the discussion here. In order to effectively bound a LQ's mass without touching its coupling to the fermions, we shall consider the physics processes where
only LQs and gauge bosons are involved. 
For example, we can measure the differential cross section of $\gamma\gamma \to \gamma\gamma$, $\gamma\gamma \to g g$ or $gg\to\gamma\gamma$. Both the SM particles and LQs contribute at loop level. The first process is the cleanest, but the first and the second ones shall only be sensitive to low-mass LQs in heavy-ion collisions. The third one might be sensitive to high-mass LQs in $p$-$p$ collisions, though it suffers from the irreducible background $q\bar{q}\to\gamma\gamma$.  
The resonance effect near the threshold of a pair of LQs shall enhance the sensitivity.
Anyway, the method presented in next section is more available and straightforward though more experimental efforts are needed.

\section{\label{sec:method} A model-independent searching method using the LQ pair production}
The LQ, $S_3$, considered in this feasibility study has a charge $|q|=\frac{4}{3}e$ and is allowed to decay to all possible $lq$ modes ($l=e,\mu,\tau$ and $q=d,s,b$). The pair production in $p$-$p$ collisions and Pb-Pb UPCs dominate if the coupling constant is not too big (e.g. $\lambda\lesssim0.1$ as shown in Fig.~\ref{fig:xs_lam}). This is consistent with the bounds from the low-energy physics in last section. The cross section of pair production does not depend on the coupling constant, but on the branching fractions of the decay modes.
Thus we consider all possible final states, namely, $eejj$, $e\mu jj$, $e\tauhad jj$, $\mu\mu jj$, $\mu\tauhad jj$, and $\tauhad\tauhad jj$, where $\tauhad$ denotes a hadronic tau decay candidate and $j$ denotes a jet candidate.  We do not attempt to identify the jet flavour for simplicity. For each final state, all LQ's contributions are considered. For example, all LQ$\to eq/\mu q/\tau q$ decays contribute in the $e\mu jj$ final state.

Three floating parameters, $\kappa, B_{ex}$ and $B_{\mu x}$, are introduced in the statistical interpretation. The branching fraction $B_{\tau x}$ is constrained to be $1-B_{ex}-B_{\mu x}$. $\kappa$ is the ratio of the measured cross section to the model prediction. We further impose the conditions $0\leq B_{ex},B_{\mu x}\leq 1$ and $B_{ex}+B_{\mu x}\leq 1$. For a given LQ mass, the measurement can determine the upper limit at 95~\% confidence level (C.L.) of $\kappa$. 
If the upper limit is less than~1, then a LQ with the mass is excluded.

The method is model-independent in the sense that it allows a LQ to couple to all possible fermions.

\section{\label{sec:MC}Cross sections  and MC simulation}
\subsection{Cross sections in UPC}\label{sec:xs_UPC}
For a nucleus $A$, the cross section for the process $A+A \to A+A+X$ is
\begin{equation}
    \sigma_{X} = \int dk_1dk_2\frac{d^2L_{\gamma\gamma}}{dk_1dk_1}\sigma^{\gamma\gamma \to X}(k_1,k_2) \:,
\end{equation}
where $\sigma^{\gamma\gamma\to X}(k_1,k_2)$ is the production cross section of the final state $X$ in two-photon interactions with energy $k_1$ and $k_2$; $d^2L_{\gamma\gamma}/dk_1dk_2$ is the corresponding two-photon luminosity.
\begin{equation}~\label{eq:lumi_yy}
    \frac{d^2L_{\gamma\gamma}}{dk_1dk_2} = \int_{b_1>R_A}\int_{b_2>R_A} d^2b_1d^2b_2\frac{d^3N_\gamma}{dk_1d^2b_1}\frac{d^3N_\gamma}{dk_2d^2b_2} \:,
\end{equation}
where $R_A$ is the nucleus radius; and $d^3N_\gamma/dkd^2b$ is the photon flux from a nucleus at a distance $b$. For a charge $Z$ nucleus with velocity $\beta$ and Lorentz boost factor $\gamma$, the photon flux is 
\begin{equation}\label{eq:photon_flux}
    \frac{d^3N_\gamma}{dkd^2b} = \frac{Z^2\alpha x^2}{\pi^2kb^2}[K_1^2(x) + \frac{K_0^2(x)}{\gamma}] \:,
\end{equation}
where $x=\frac{kb}{\beta\gamma}$ and $K_0(x)$ and $K_1(x)$ are modified Bessel functions. The photon flux decreases exponentially above the cut off $k_{\text{max}}\approx \frac{\gamma}{R_A}$. 

In Eq.~\ref{eq:lumi_yy}, we have assumed that the probability of not having a hadronic interaction, denoted by $P_{0\text{had}}(b)$, is 0 if $b<R_A$ and 1 if $b>R_A$. In general, we can replace the integration $\int_{b>R_A}\cdots d^2b$ in Eq.~\ref{eq:lumi_yy} by $\int_{b=0}^\infty P_{0\text{had}}(b)\cdots d^2b$. Using a more complicated description of $P_{0\text{had}}(b)$ leads to a $10-15$\% difference in the flux~\cite{UPCreview1}. 

\subsection{MC sample preparation}
The MadGraph~5~\cite{MG5} is used to compute cross sections by convolving the Weiz\"acker-Williams equivalent photon flux (Eq.~\ref{eq:photon_flux}) of the nucleus with the elementary two-photon cross section, and to produce the signal and background samples. The parton showers are simulated with Pythia~8~\cite{pythia} and the detector response is simulated with Delphes~3~\cite{delphes} based on the Run~2 performance of the ATLAS detector.

The signal model is obtained from~\cite{LQtoolboxurl} and documented in Ref.~\cite{LQtoolbox}. The LQ of interest in this work is the scalar boson $S_3$, which is allowed to couple to all charged leptons and down-type quarks. 
Six signal samples corresponding to the processes $\gamma\gamma \to S_3\bar{S}_3 \to eeqq/e\mu qq/e\tau qq/\mu\mu qq/\mu\tau qq/\tau\tau qq$ are produced with a mass 100~GeV and all coupling constant fixed at $10^{-2}$. The possible interference effect with the SM process is not considered. 
The fiducial region at generator level is defined as $\pt(l)>13$~GeV, $\pt(j)>15$~GeV, $|\eta(l)|<2.5$, $|\eta(j)|<5$, $\Delta R(j,l)>0.3$ and $\Delta R(j,j)>0.3$, where $\pt$, $\eta$ and
$\Delta R$ represent the transverse momentum, rapidity and the angular distance\footnote{The angular distance between two objects is defined as $\Delta R(a,b)\equiv \sqrt{(\eta(a)-\eta(b))^2+(\phi(a)-\phi(b))^2}$, where $\eta$ and $\phi$ are the rapidity and azimuthal angle, respectively.}, respectively.

\begin{figure}[htbp]
    \centering
    \includegraphics[width=0.45\textwidth]{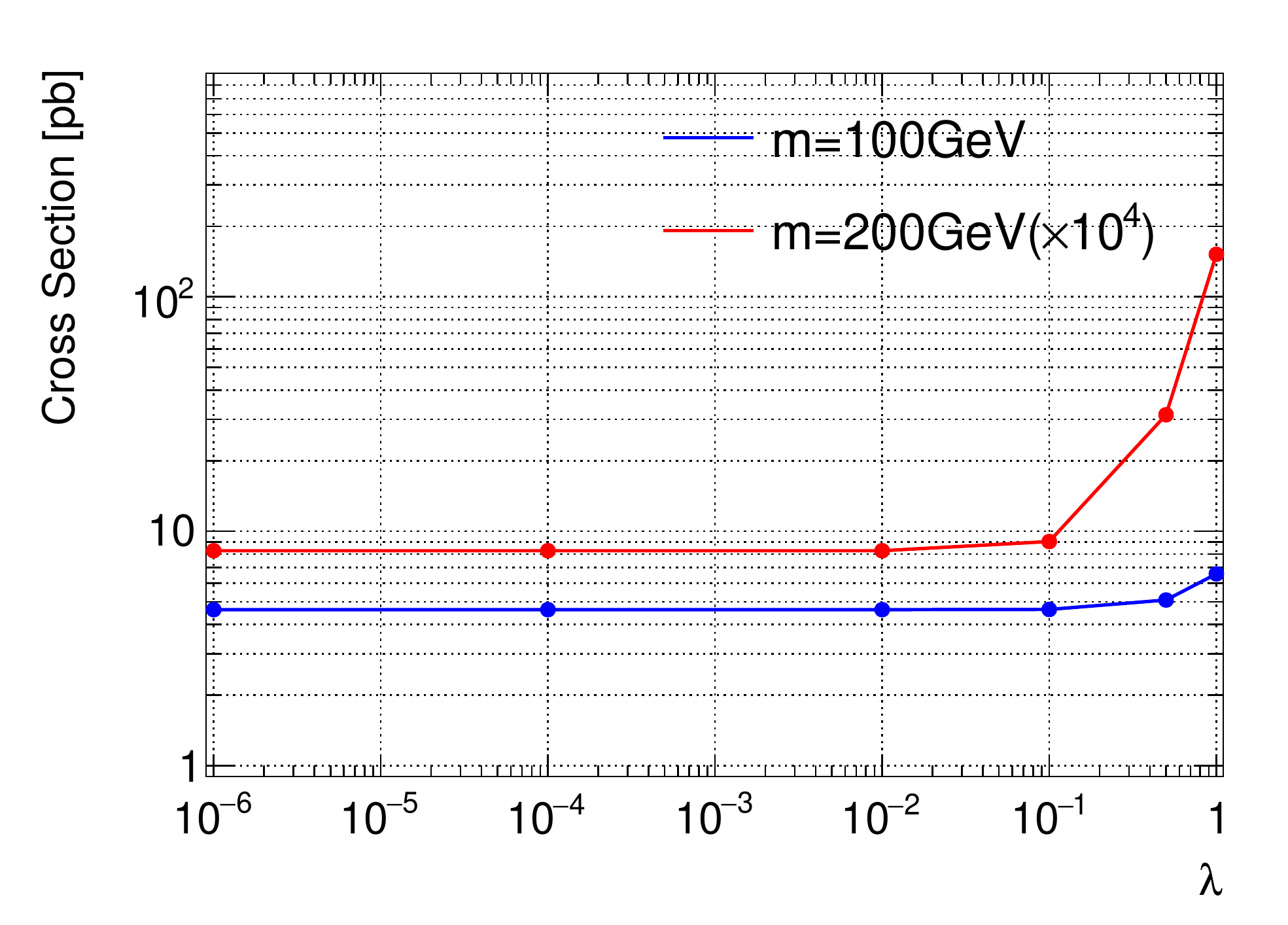}
    \caption{\label{fig:xs_lam}
    Fiducial cross section of Pb$+$Pb$\to$Pb$+$Pb$+e^+e^-d\bar{d}$ driven by the leptoquark $S_3$ in ultra-peripheral collisions at $\sqrt{s}=5.03$~TeV as a function of the coupling constant $\lambda_{ed}$ with a mass of 100~GeV (blue curve) or 200~GeV (red curve, scaled by a factor of $10^4$). 
    }
\end{figure}

Figure~\ref{fig:xs_lam} shows the fiducial cross section of Pb$+$Pb$\to$Pb$+$Pb$+e^+e^-d\bar{d}$ as a function of the coupling constant $\lambda_{ed}$. 
Little dependence is seen for $\lambda_{ed}\lesssim0.1$. The cross section used in the normalization does not consider the branching fractions, which will be treated as the parameters of interest in the statistical interpretation, and is scaled by a factor of $10^{-2}$ in the plots in Sec.~\ref{sec:selection} just for better illustration.

The dominant background is from the SM process $\gamma\gamma \to l^+l^-q\bar{q}$. It is irreducible. The background events with multiple leptons or multiple jets are also considered and their contribution is minor. In addition, the background from Pb-Pb hadronic interactions is inevitable.  It is very difficult to simulate this background in reality. We usually design tight cuts to suppress it and estimate it in a data-driven way. We assume this background is negligible after the event selection in next section.   

\section{Event selection and Observable reconstruction}\label{sec:selection}
Full reconstruction of all 4 objects is not desirable because of low efficiency. We require two lepton candidates and at least one jet candidates to be reconstructed.
A charged lepton candidate ($e/\mu$) is required to have transverse momentum $\pt>15$~GeV and rapidity $|\eta|<2.5$. A hadronic tau candidate $\tauhad$ is required to have $\pt>20$~GeV and $|\eta|<2.5$. The associated number of charged tracks is required to be $1$ or $3$ with unit sum of charges. 
A jet candidate denoted by $j$  is required to have $\pt>20$~GeV and $|\eta|<5.0$. 

An overlapping removal is performed according to the priority order $\mu\to e\to\tauhad \to j$. We first select muon candidate and then select electron candidate. The electron candidate is required not to overlap with the muon candidate within the angular distance $\Delta R=0.4$. $\tauhad$ candidate is required not to overlap with electron and muon candidates within the same angular distance. Similarly the $j$ candidate is required not to overlap with $\mu$, $e$ and $\tauhad$ candidates. 
12 signal regions (SR) are designed below ($x=1,2$).
\begin{itemize}
    \item ``ee$x$j'' signal region: 2 electron candidates with opposite charge sign and $x$ jets.
    \item ``emu$x$j'' signal region: 1 electron candidate and 1 muon candidate with opposite charge sign and $x$ jets.
    \item ``eta$x$j'' signal region: 1 electron candidate and 1 $\tauhad$ candidates with opposite charge sign and $x$ jets.
    \item ``mumu$x$j'' signal region: 2 muon candidates with opposite charge sign and $x$ jets.
    \item ``muta$x$j'' signal region: 1 muon candidate and 1 $\tauhad$ candidate with opposite charge sign and $x$ jets.
    \item ``tata$x$j'' signal region: 2 $\tauhad$ candidate and $x$ jets. For the signal region ``tata2j'', the missing transverse energy~\footnote{The missing transverse energy is defined as negative vectorial sum of all visible objects' transverse momenta.} $\etmiss$ is required to be greater than 10~GeV to suppress the multi-jet background.
\end{itemize}
If we have real data samples, we can apply further requirements to suppress the background due to the hadronic Pb-Pb interactions: 1) no additional charged tracks in the inner detector and no additional energy deposit from neutral particles in the calorimeters as used in Ref.~\cite{exclWW,yy2WW}; 2) no significant energy deposit in the Zero Degree Calorimeters~\cite{ZDC} for the ATLAS detector ( because the Pb nuclei are intact in UPCs) as
used in Ref.~\cite{yy2mumu}; 3) adoption of multi-variate-analysis (MVA) methods~\cite{TMVA,weaklearner} to separate signal events from background events. In Appendix~\ref{app:purity}, it is demonstrated that the signal purity in Pb-Pb collisions is much better than that in $p$-$p$ collisions.

The signal is identified as a peak structure in the invariant mass spectrum of a lepton and a jet candidates. For the events with only one jet, the jet is used twice to reconstruct the LQ mass. For the events with two jets, the best alignment is determined by minimizing their relative mass difference, namely,
\begin{equation}
    \min \frac{|m(l_1j_1)-m(l_2j_2)|}{m(l_1j_1)+m(l_2j_2)} \:.
\end{equation}
The subscript associated with the lepton candidate represents the charge sign, namely, $l_{1,2}=l^{\pm}$, in the signal regions with the same flavour leptons (ee$x$j, mumu$x$j and tata$x$j). Otherwise the subscript represents the lepton flavour. For the eta$x$j and muta$x$j signal regions, we assume that the missing transverse energy is due to the neutrinos from the tau decays and thus added to the $\tauhad$ candidate, namely,
\begin{eqnarray}
    p_x(\tau) &=& p_x(\tauhad) + \etmiss\cos\phimiss \:, \\
    p_y(\tau) &=& p_y(\tauhad) + \etmiss\sin\phimiss \:, \\
    p_z(\tau) &=& p_z(\tauhad) + \etmiss\sinh\eta(\tauhad) \:, \label{eq:pz_coll}
\end{eqnarray}
where $\phimiss$ is the azimuthal angle of the missing transverse energy; and the collinear approximation, which assumes that the flight direction of the invisible neutrino(s) in a tau decay is the same as that of the visible tau decay products ~\cite{mmc,mxlg}, is used in Eq.~\ref{eq:pz_coll}. 

\begin{figure}[htbp]
    \centering
    \includegraphics[width=0.35\textwidth]{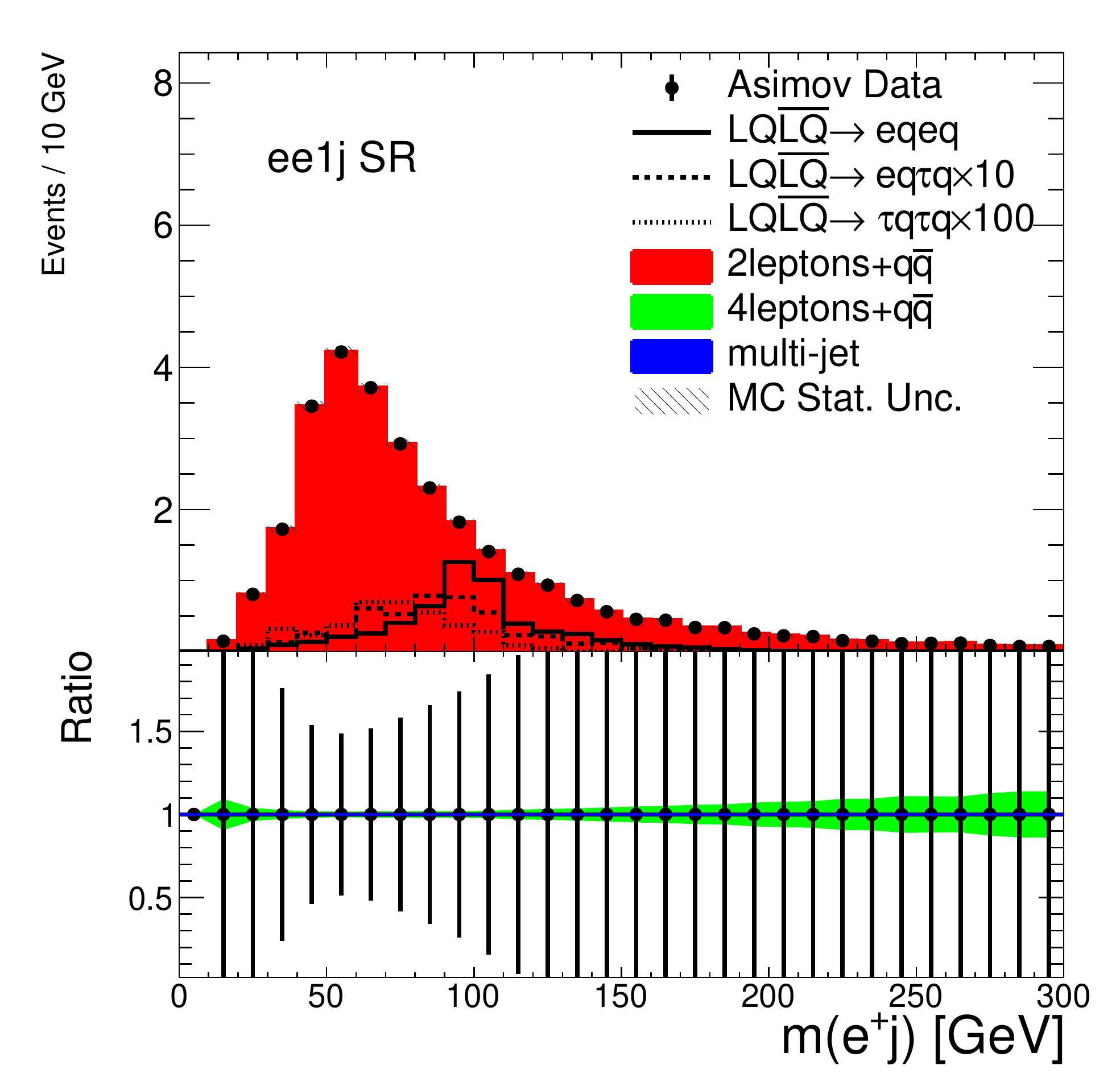}
    \includegraphics[width=0.35\textwidth]{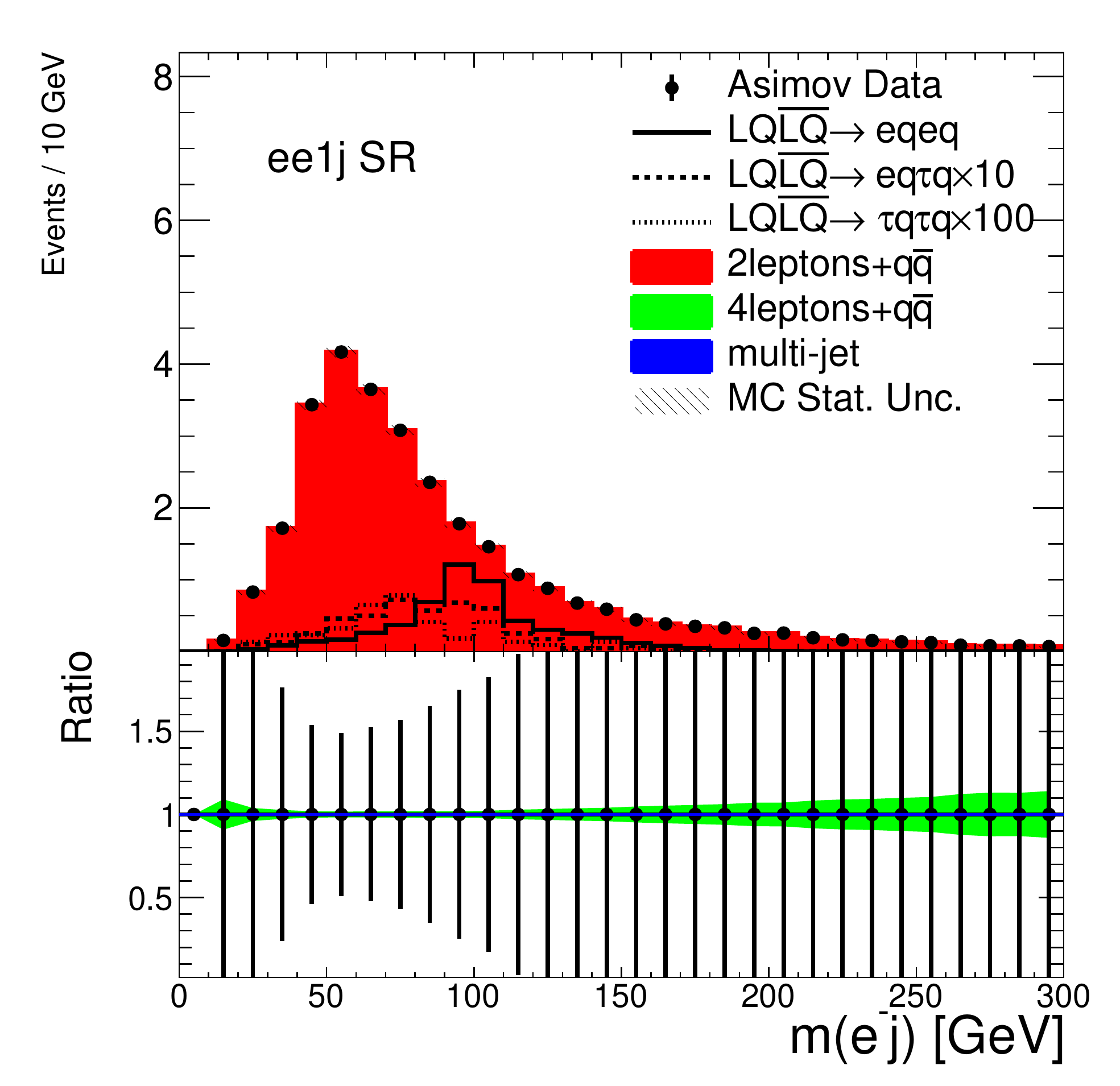}\\
    \includegraphics[width=0.35\textwidth]{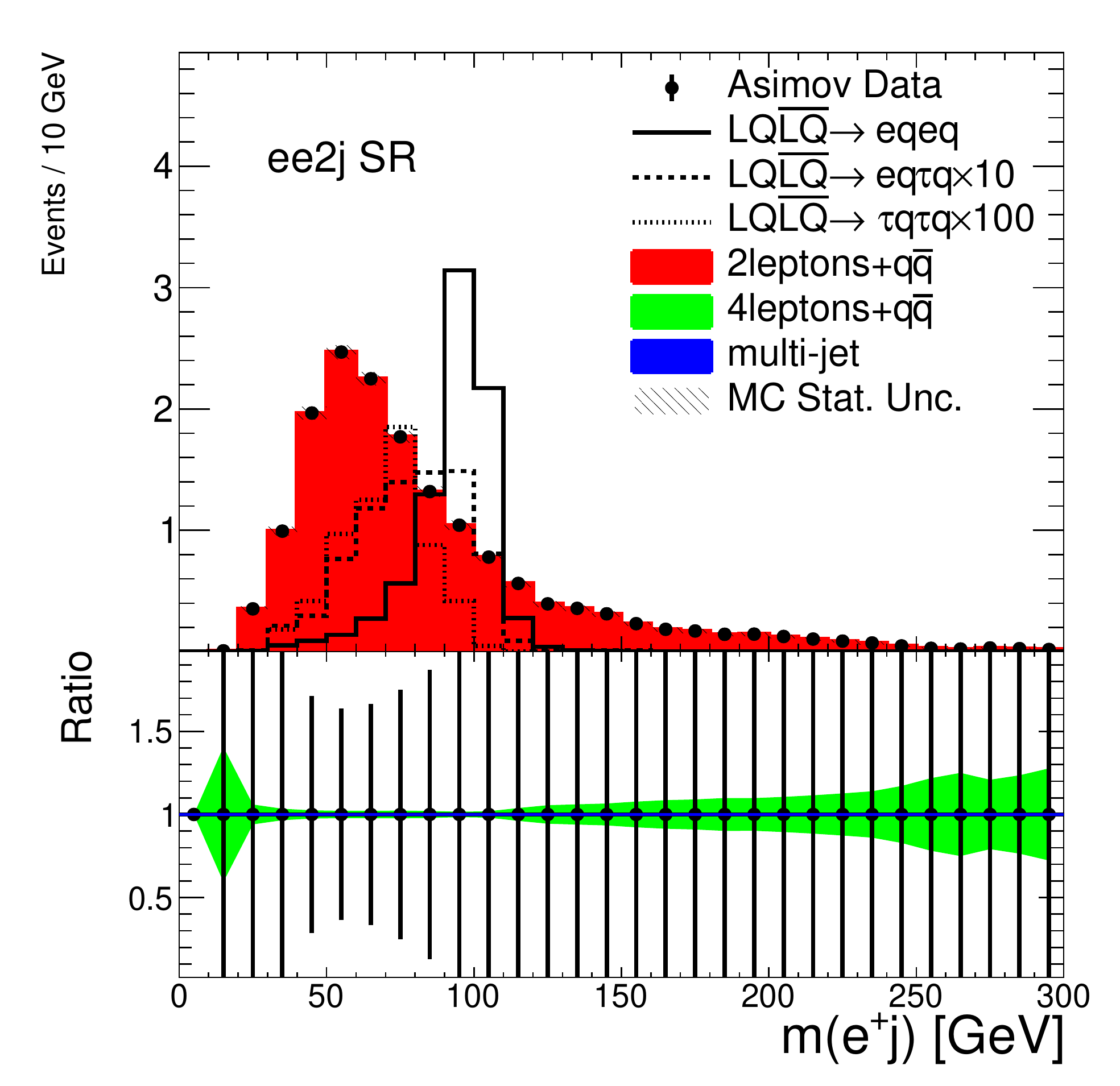}
    \includegraphics[width=0.35\textwidth]{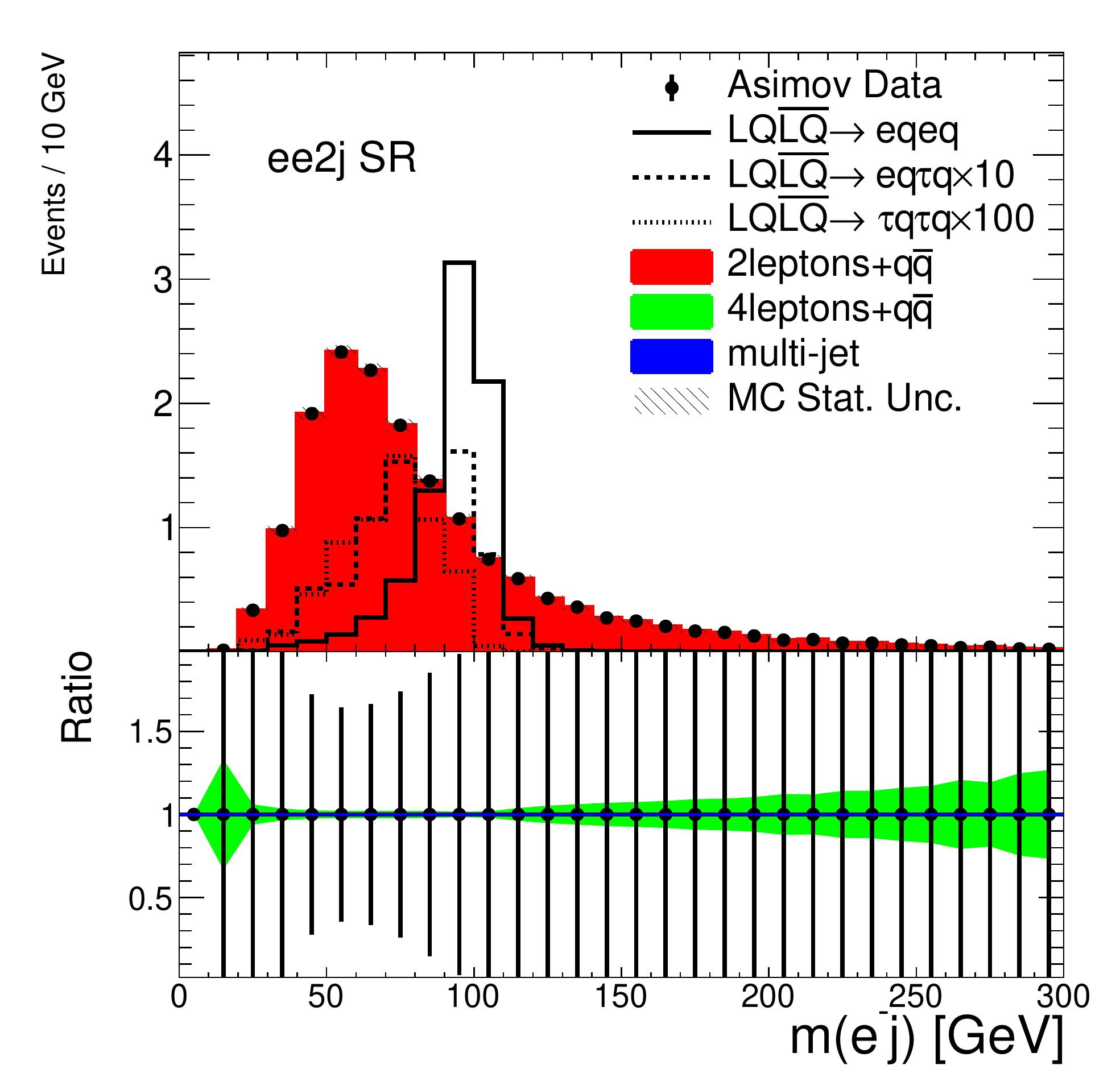}
    \caption{
        \label{fig:m_lq_ee2j}
        The distribution of the invariant mass of a lepton candidate and a jet candidate in the signal region ``ee1j'' (Top) and ``ee2j'' (Bottom).
        The left one is $m(e^+j)$ and the right one is $m(e^-j)$.
        The black points in the upper pads show the asimov data which is just the sum of all background events with uncertainties completely suppressed for better illustration. The ratio of the asimov data and the total background is shown in the lower pads, where the error bars on the black points represent the expected data uncertainty while the green bands represent the total MC statistical uncertainty. 
    }
\end{figure}
\begin{figure}[htbp]
    \centering
    \includegraphics[width=0.35\textwidth]{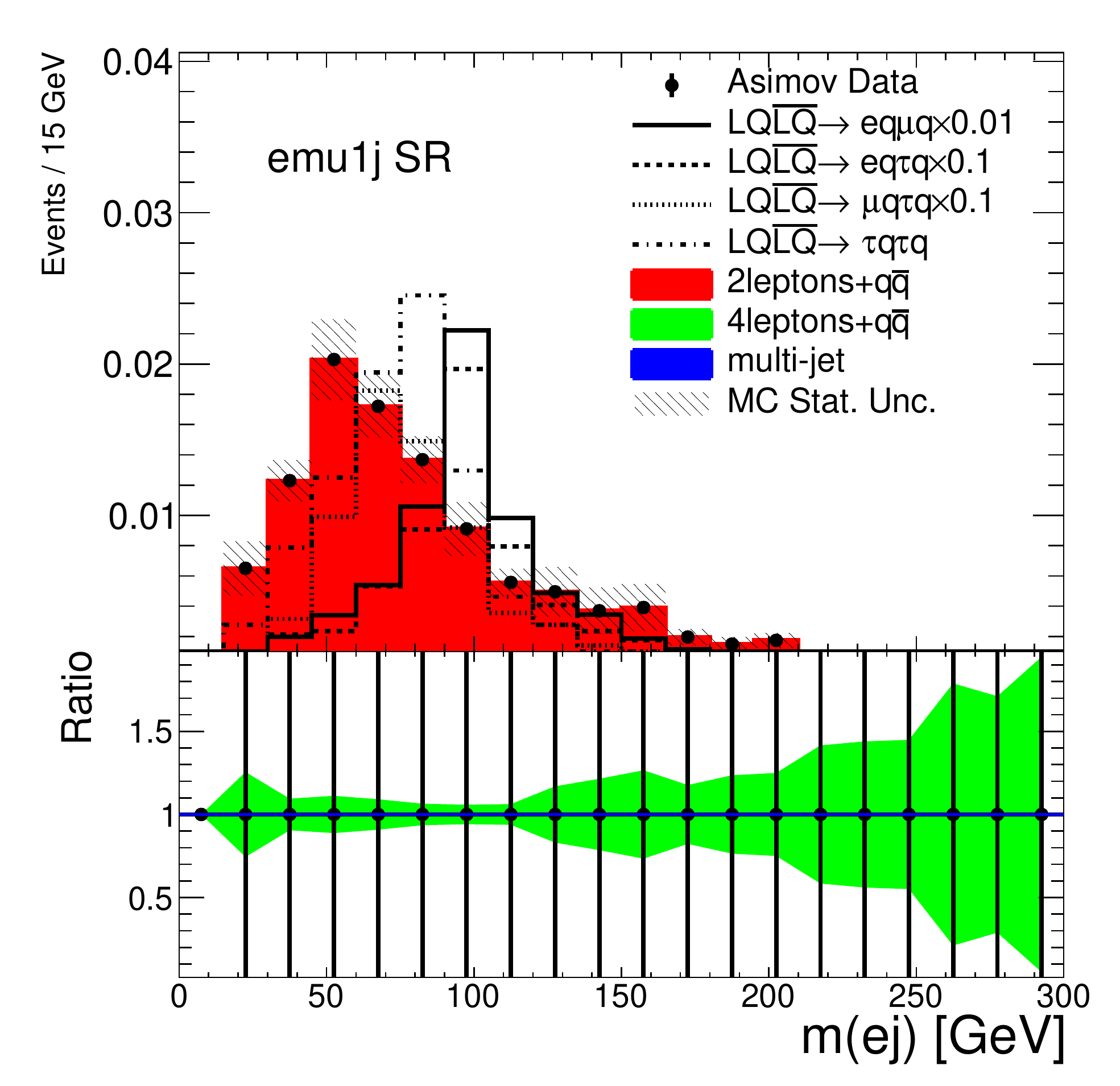}
    \includegraphics[width=0.35\textwidth]{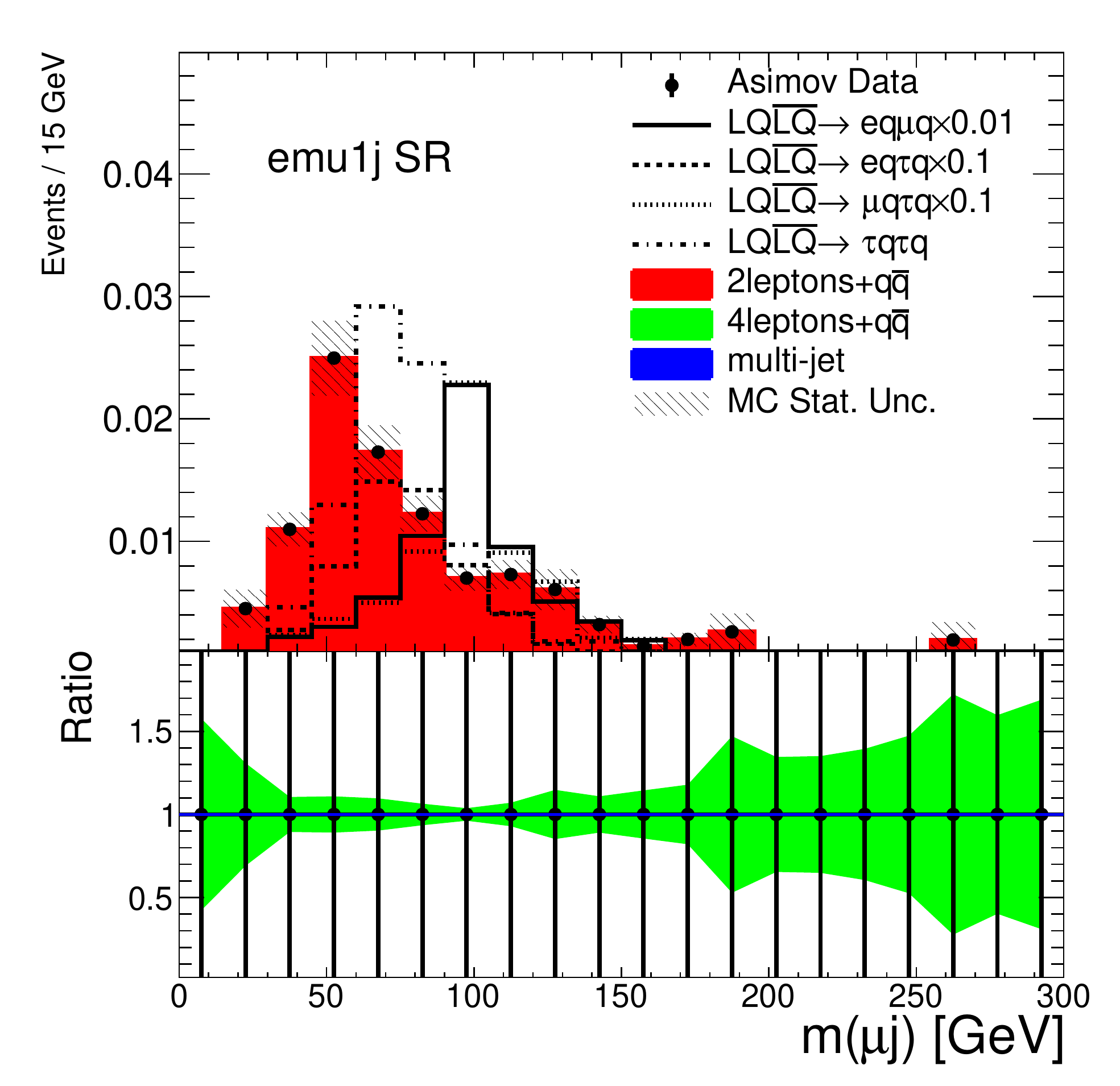}\\
    \includegraphics[width=0.35\textwidth]{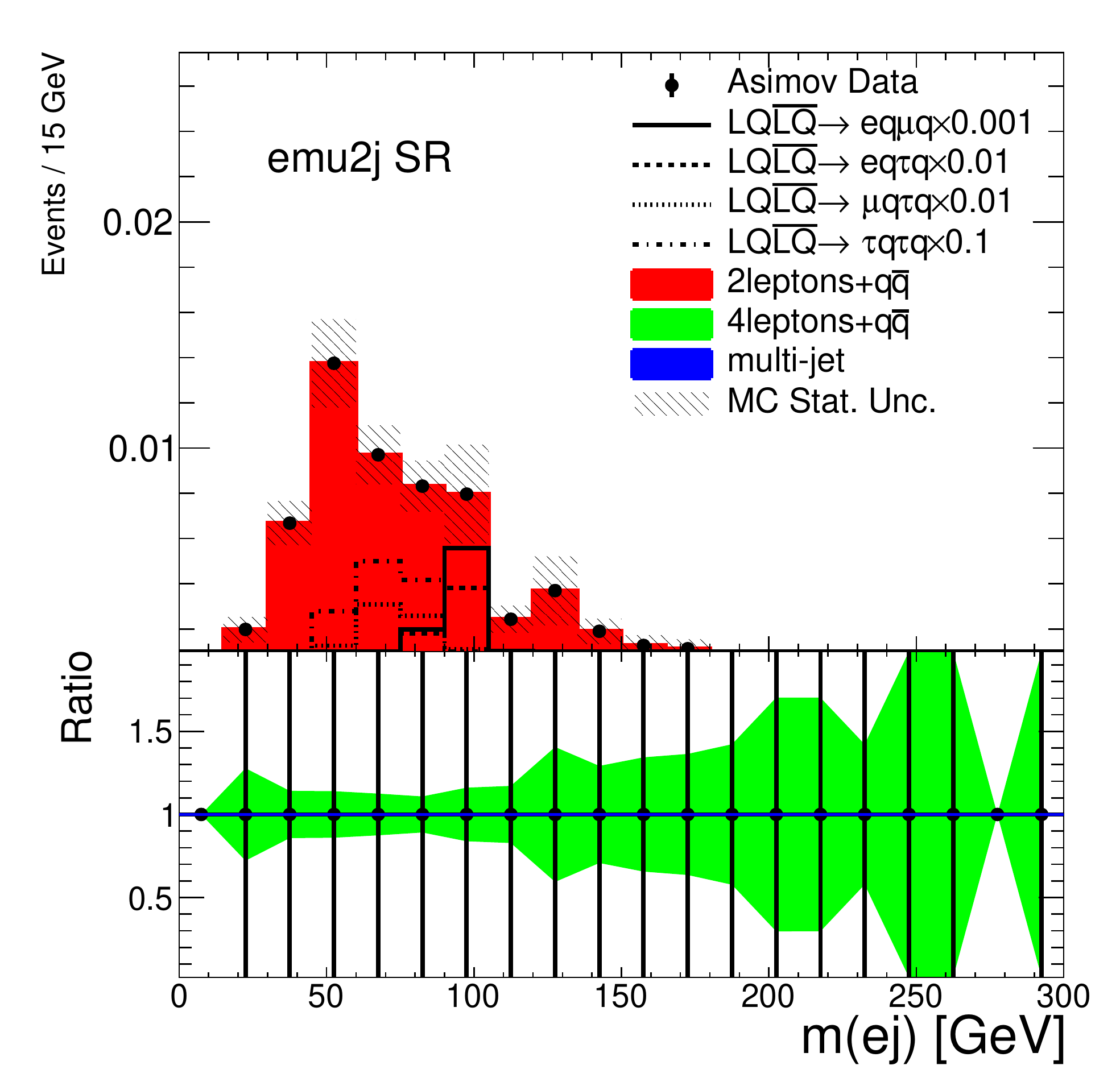}
    \includegraphics[width=0.35\textwidth]{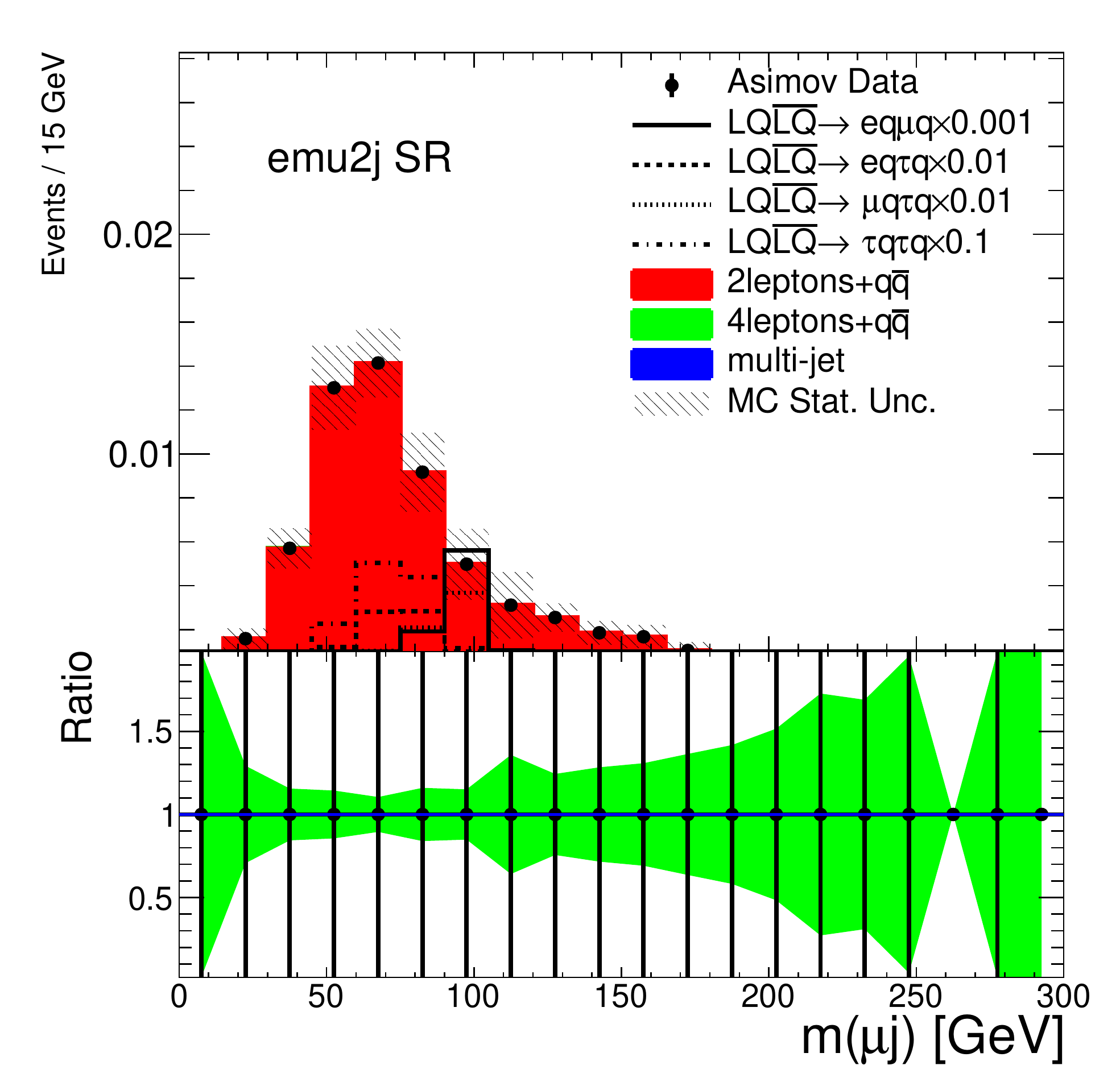}
    \caption{
        \label{fig:m_lq_emu2j}
        The distribution of the invariant mass of a lepton candidate and a jet candidate in the signal region ``emu1j'' (Top) and ``emu2j'' (Bottom).
        The left one is $m(ej)$ and the right one is $m(\mu j)$.
        The black points in the upper pads show the asimov data which is just the sum of all background events with uncertainties completely suppressed for better illustration. The ratio of the asimov data and the total background is shown in the lower pads, where the error bars on the black points represent the expected data uncertainty while the green bands represent the total MC statistical uncertainty. 
    }
\end{figure}
\begin{figure}[htbp]
    \centering
    \includegraphics[width=0.35\textwidth]{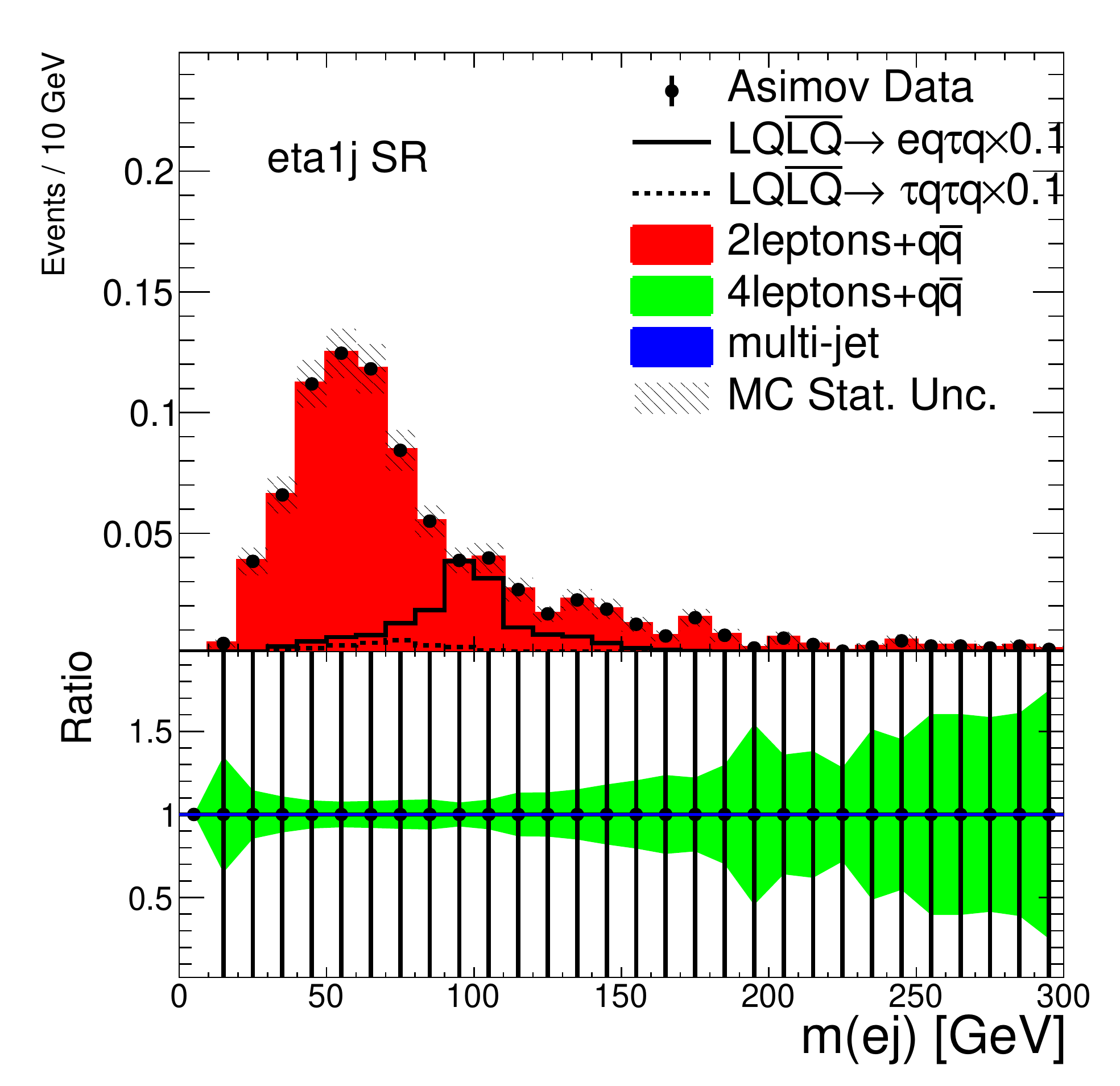}
    \includegraphics[width=0.35\textwidth]{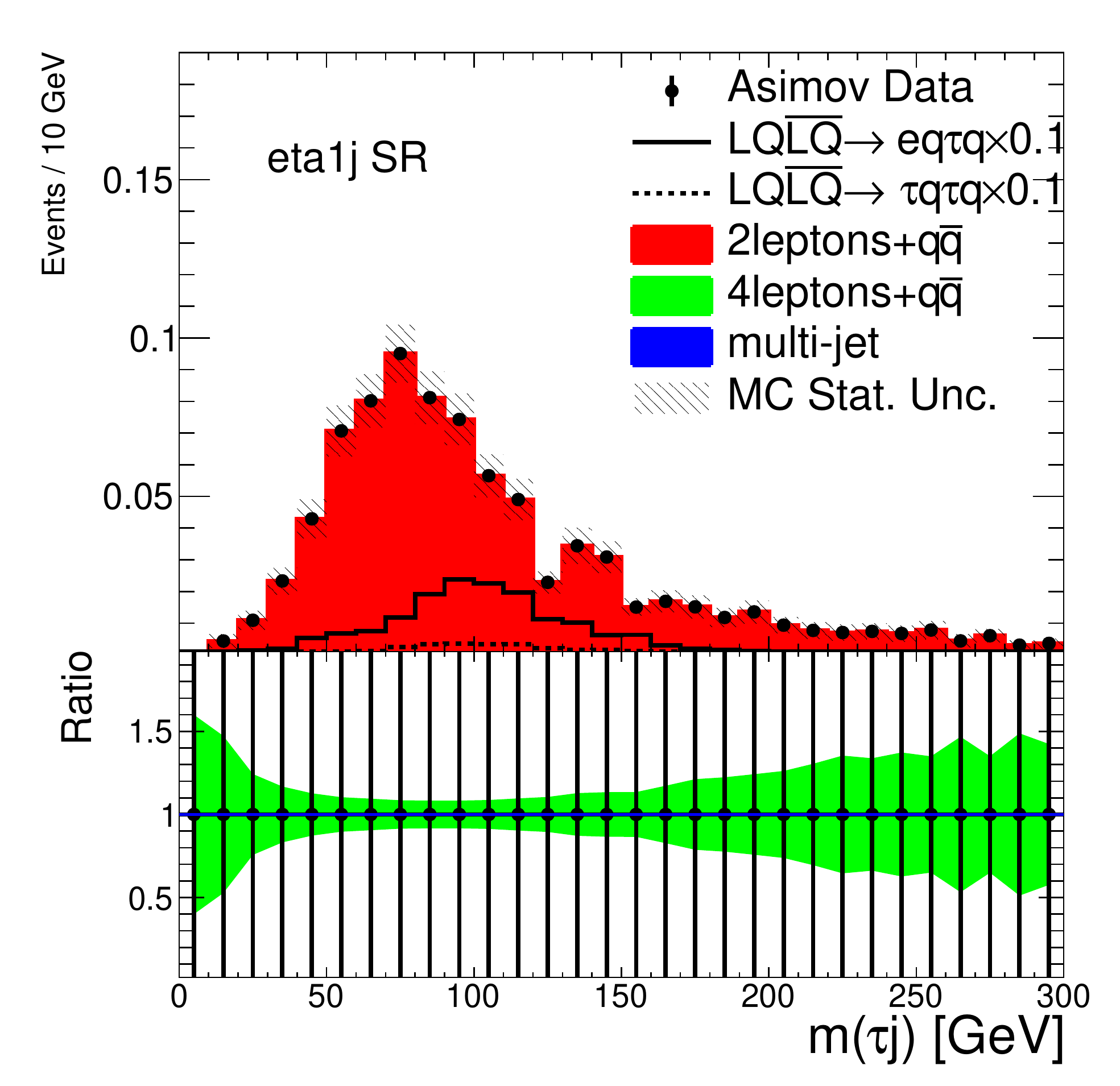}\\
    \includegraphics[width=0.35\textwidth]{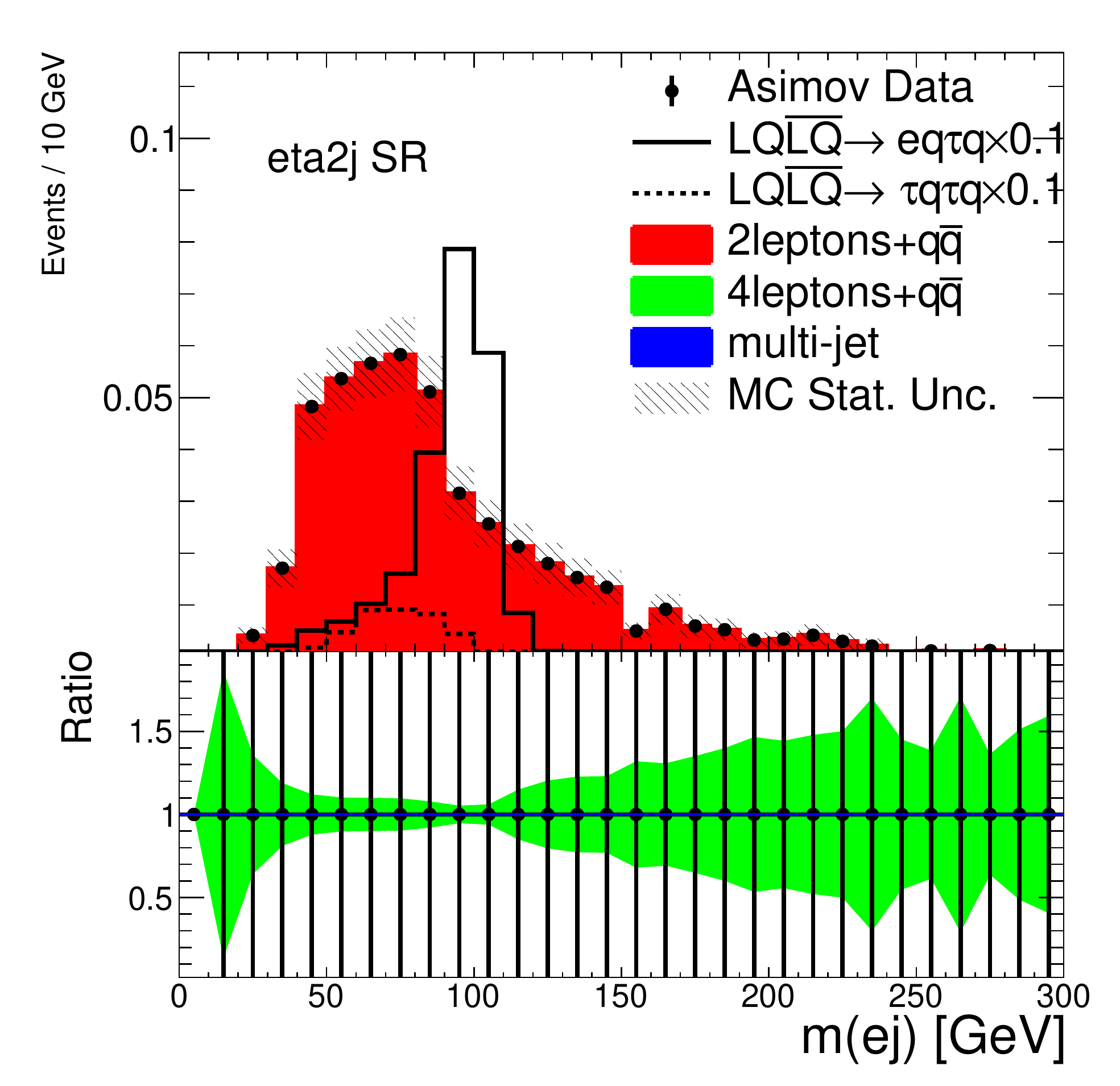}
    \includegraphics[width=0.35\textwidth]{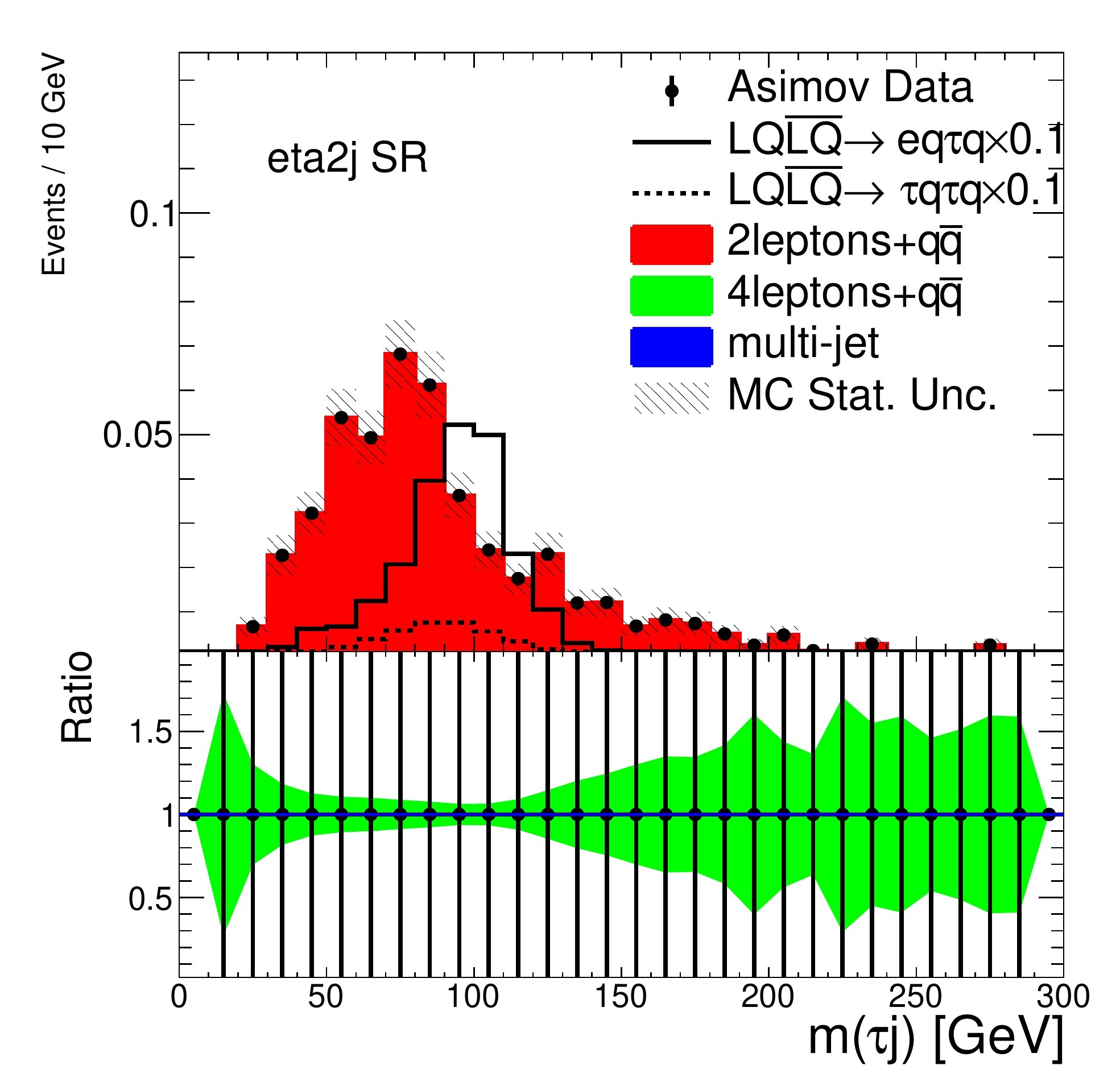}
    \caption{\label{fig:m_lq_eta2j}
        The distribution of the invariant mass of a lepton candidate and a jet candidate in the signal region ``eta1j''(Top) and ``eta2j''(Bottom).
        The left one is $m(ej)$ and the right one is $m(\tau j)$.
        The black points in the upper pads show the asimov data which is just the sum of all background events with uncertainties completely suppressed for better illustration. The ratio of the asimov data and the total background is shown in the lower pads, where the error bars on the black points represent the expected data uncertainty while the green bands represent the total MC statistical uncertainty. 
    }
\end{figure}
\begin{figure}[htbp]
    \centering
    \includegraphics[width=0.35\textwidth]{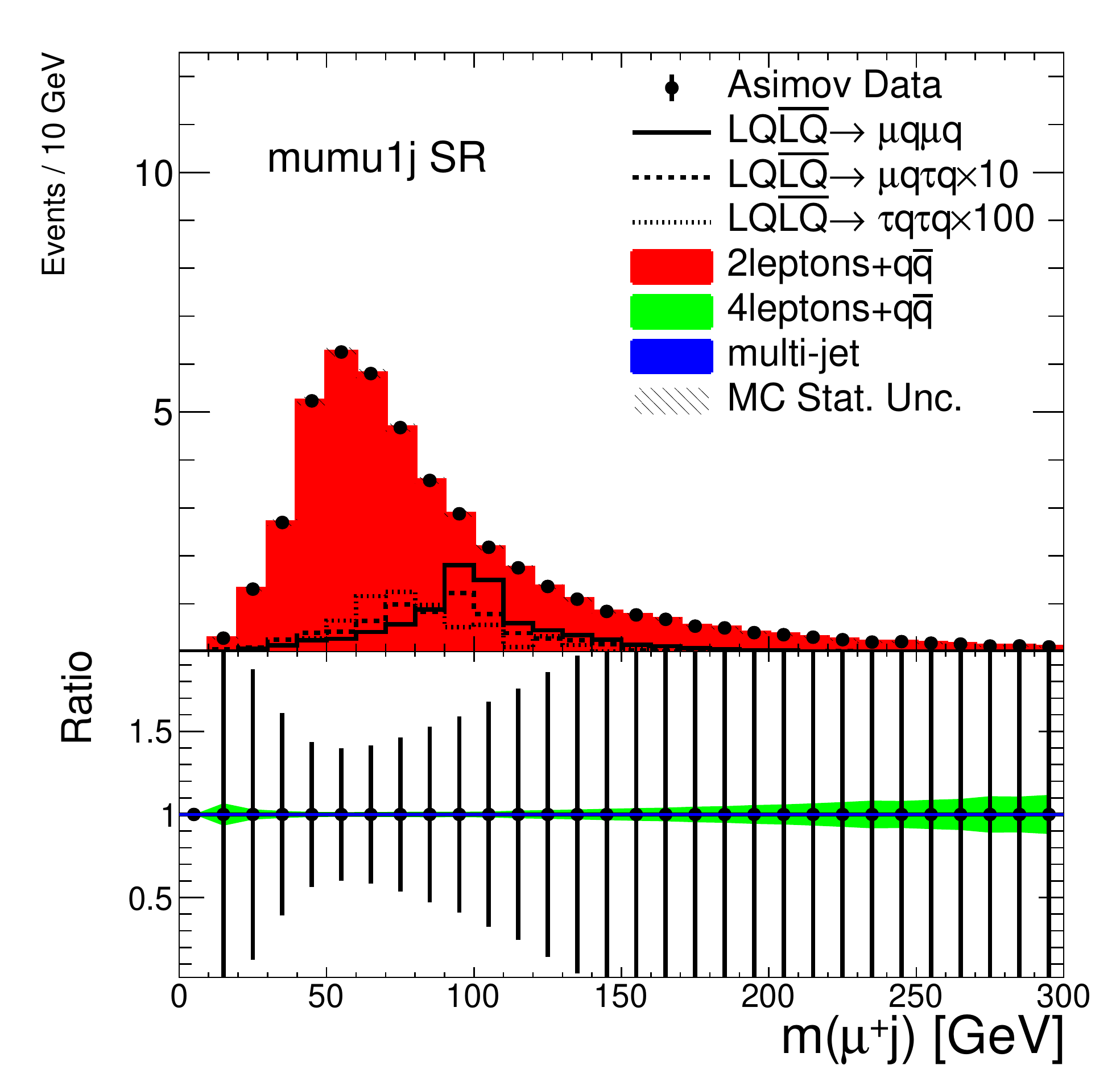}
    \includegraphics[width=0.35\textwidth]{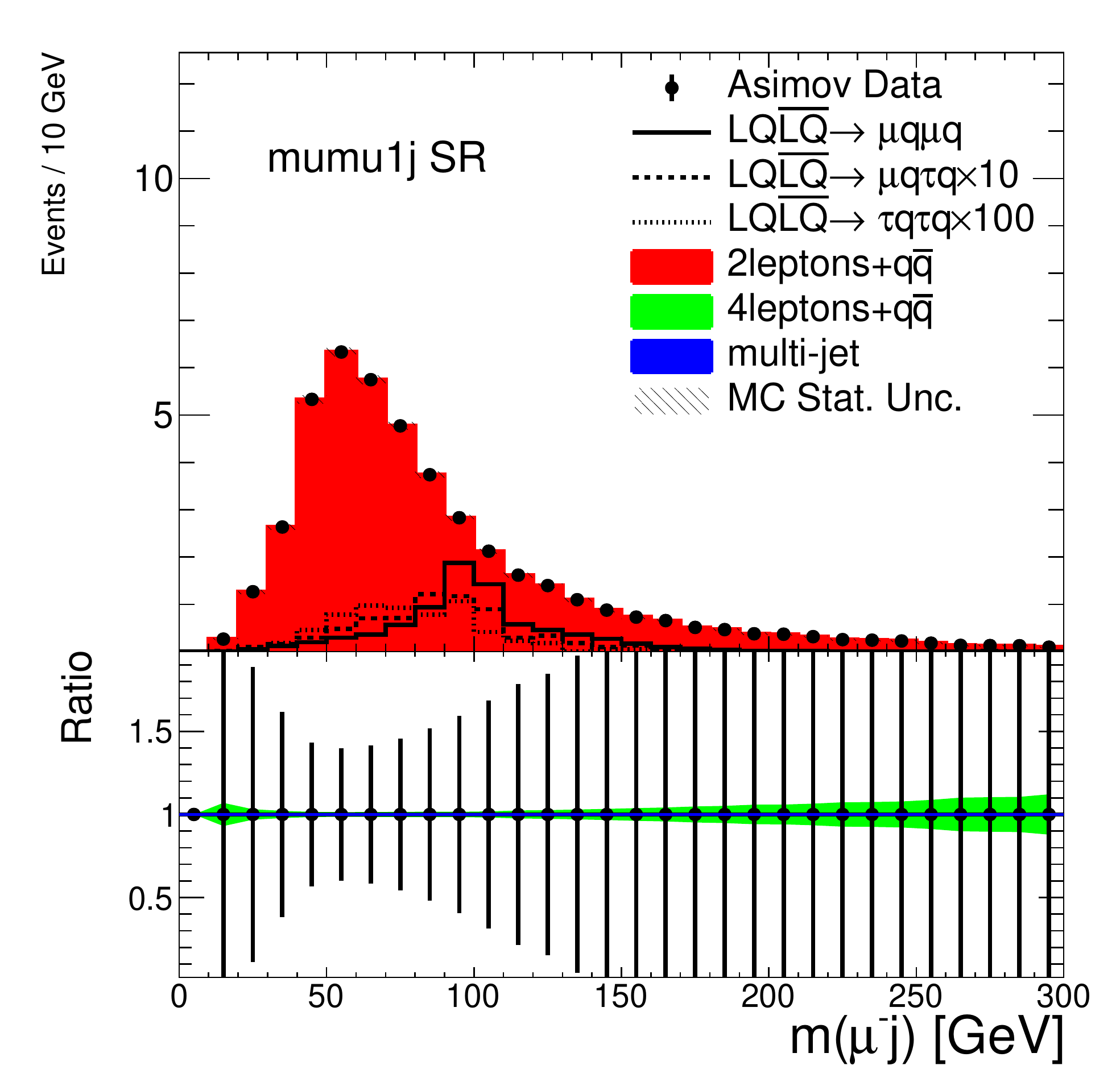}\\
    \includegraphics[width=0.35\textwidth]{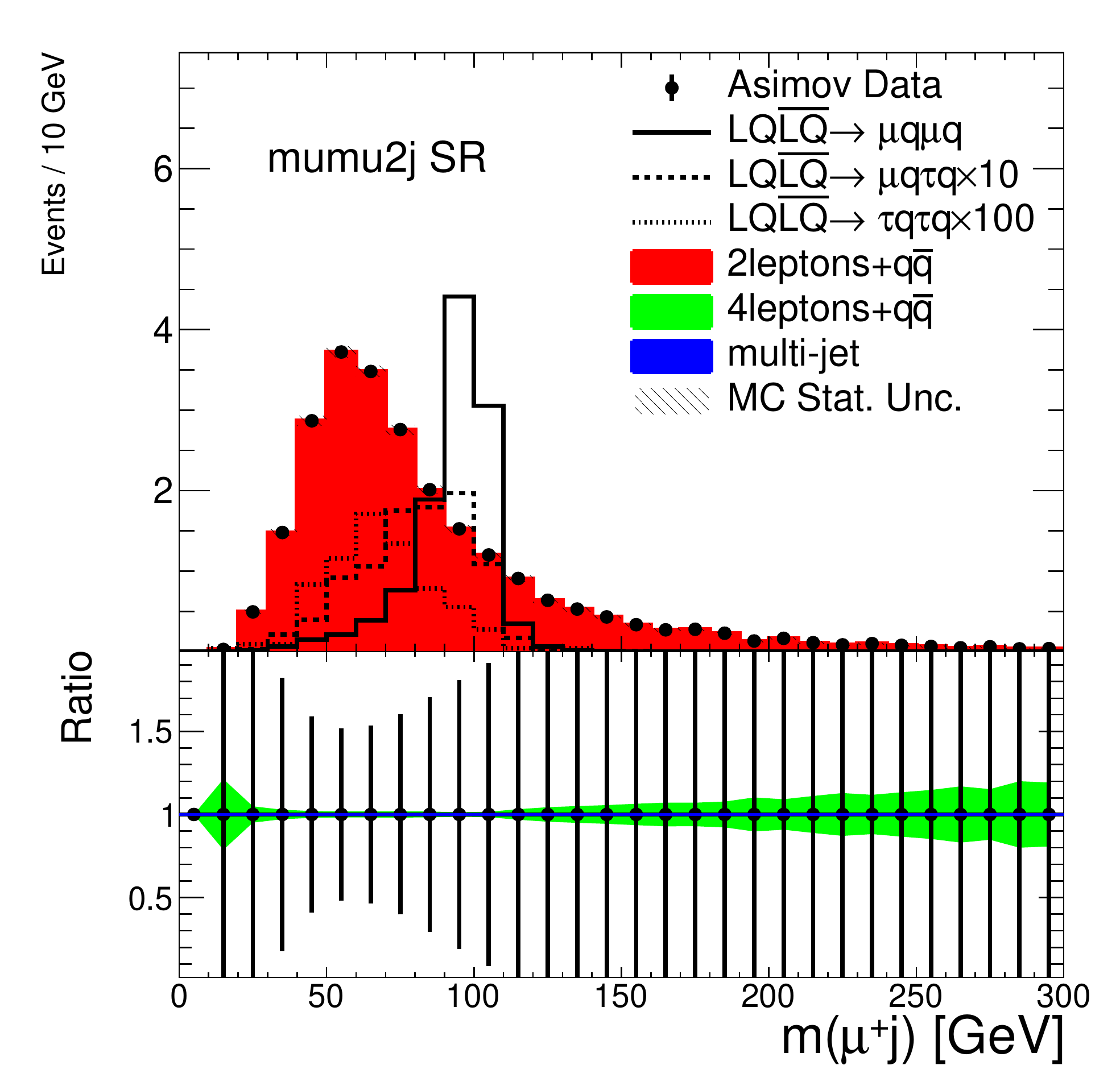}
    \includegraphics[width=0.35\textwidth]{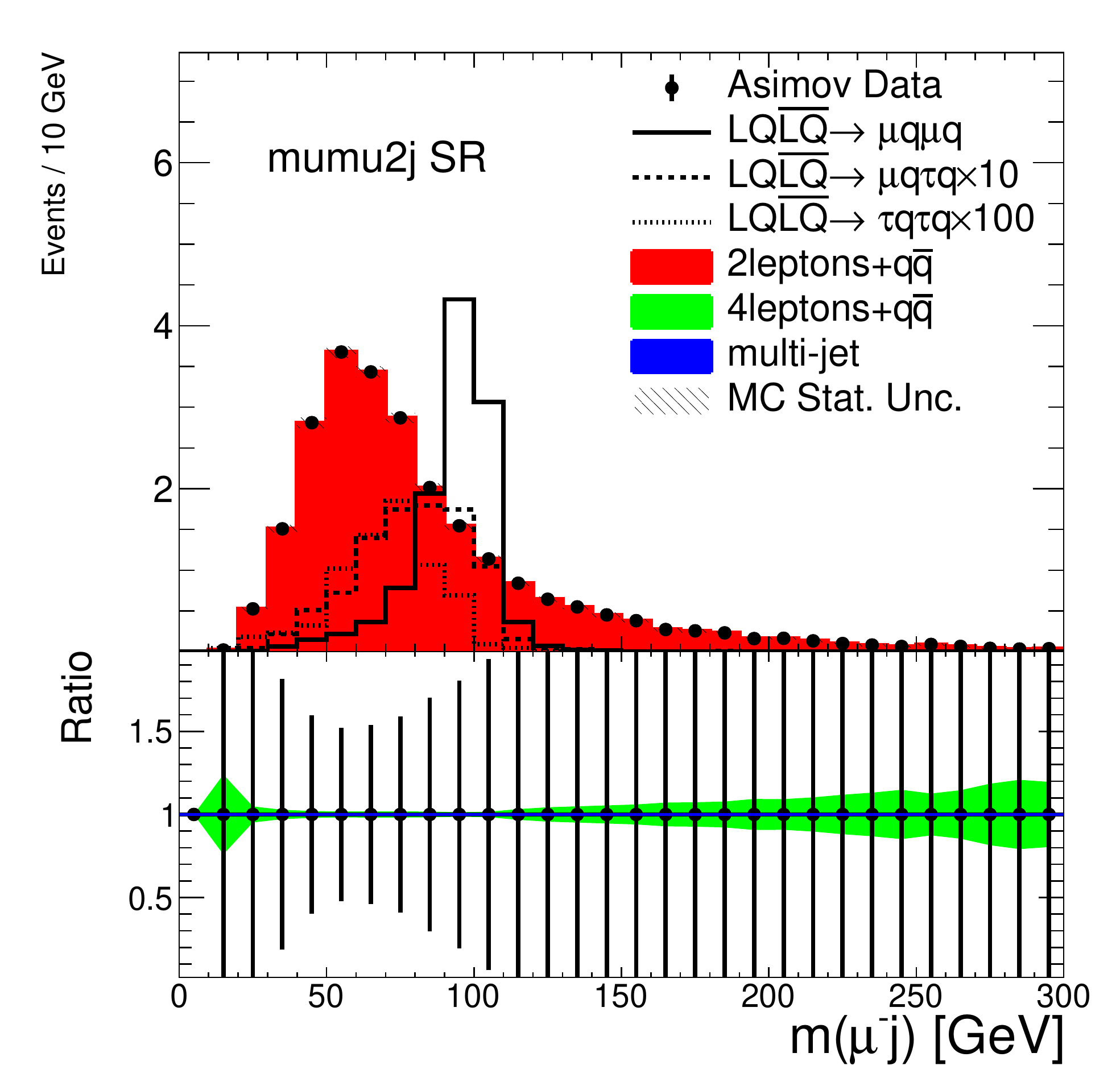}
    \caption{\label{fig:m_lq_mumu2j}
        The distribution of the invariant mass of a lepton candidate and a jet candidate in the signal region ``mumu1j''(Top) and ``mumu2j''(Bottom).
        The left one is $m(\mu^+j)$ and the right one is $m(\mu^-j)$.
        The black points in the upper pads show the asimov data which is just the sum of all background events with uncertainties completely suppressed for better illustration. The ratio of the asimov data and the total background is shown in the lower pads, where the error bars on the black points represent the expected data uncertainty while the green bands represent the total MC statistical uncertainty. 
    }
\end{figure}
\begin{figure}[htbp]
    \centering
    \includegraphics[width=0.35\textwidth]{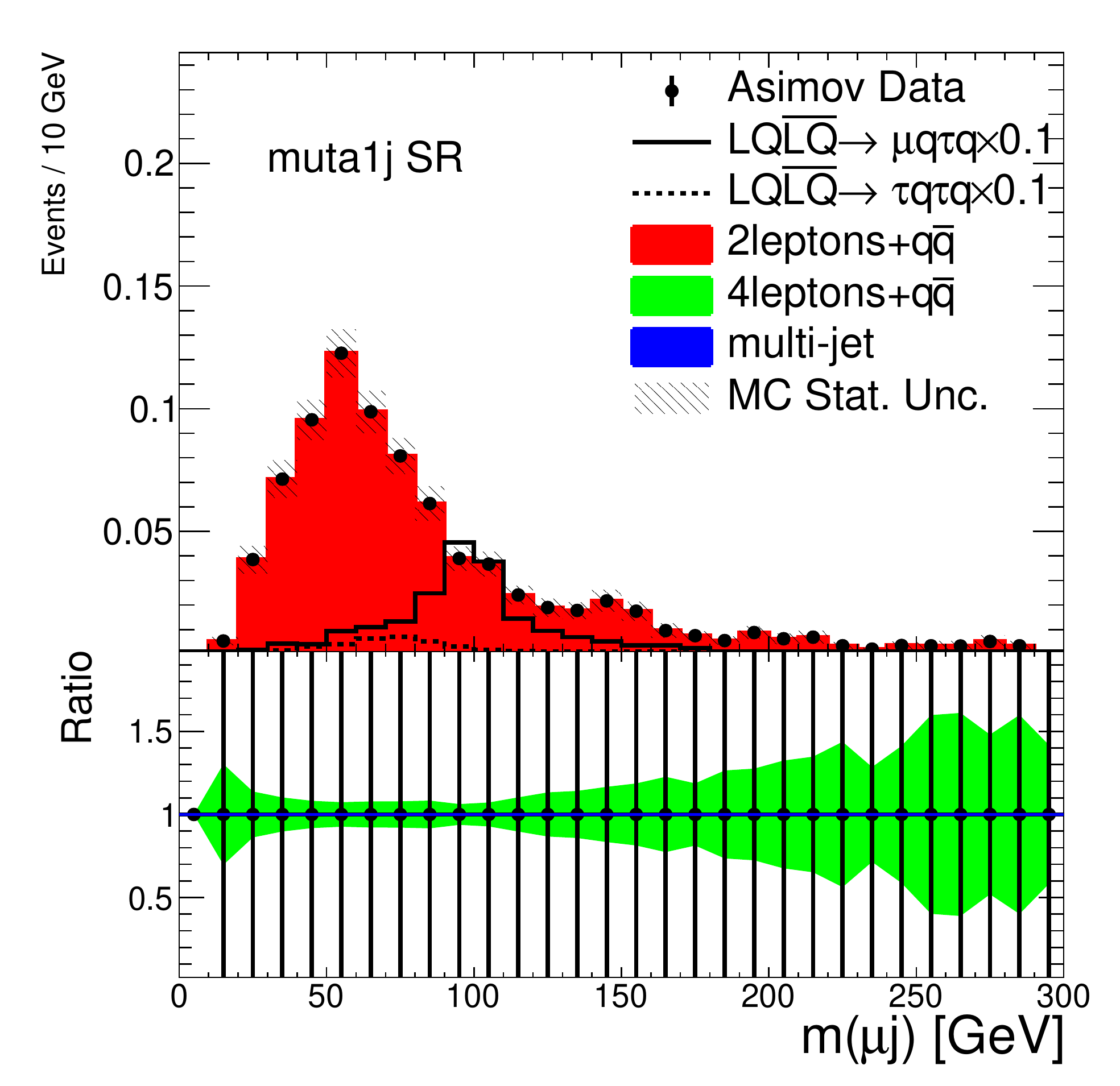}
    \includegraphics[width=0.35\textwidth]{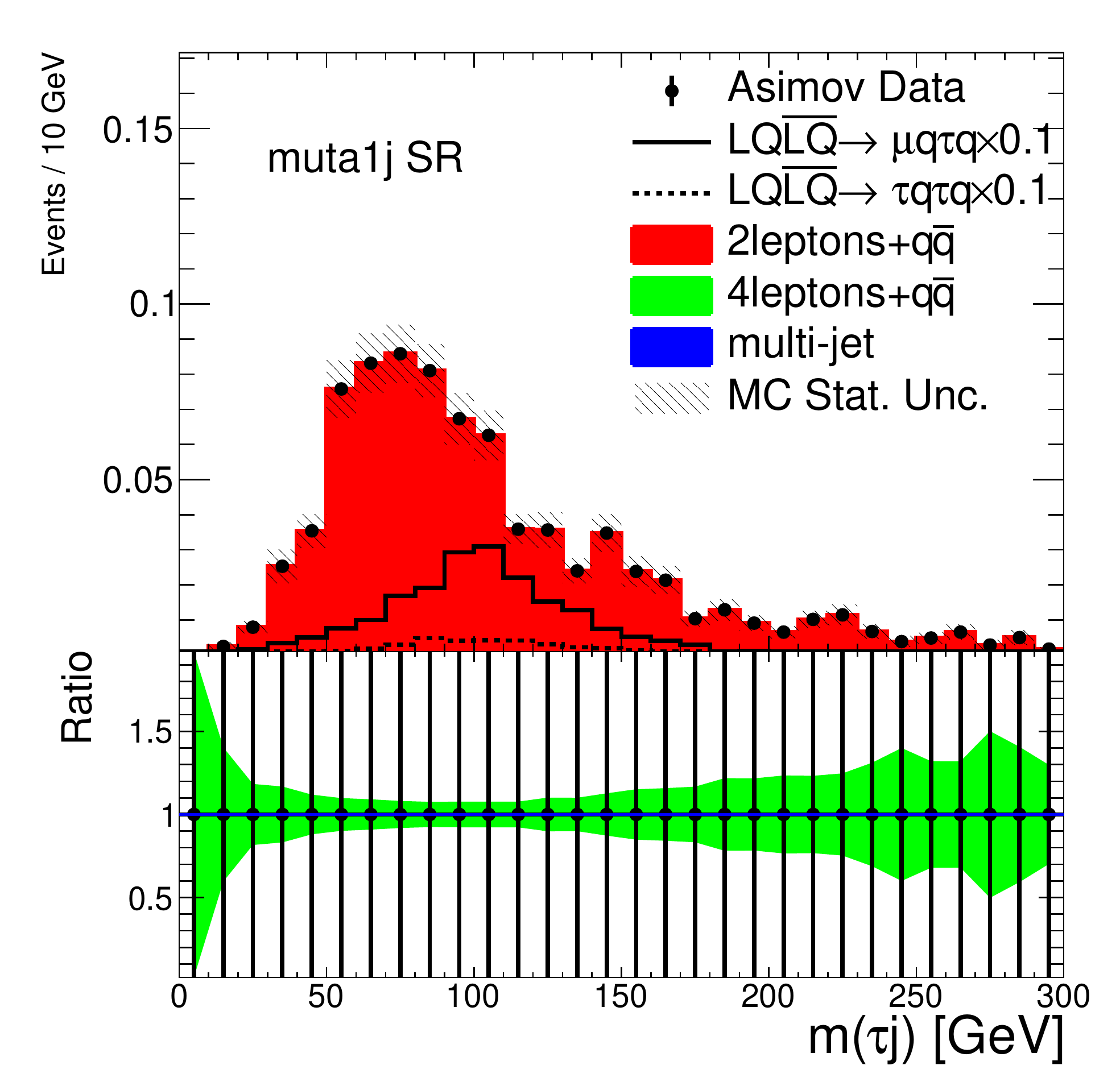}\\
    \includegraphics[width=0.35\textwidth]{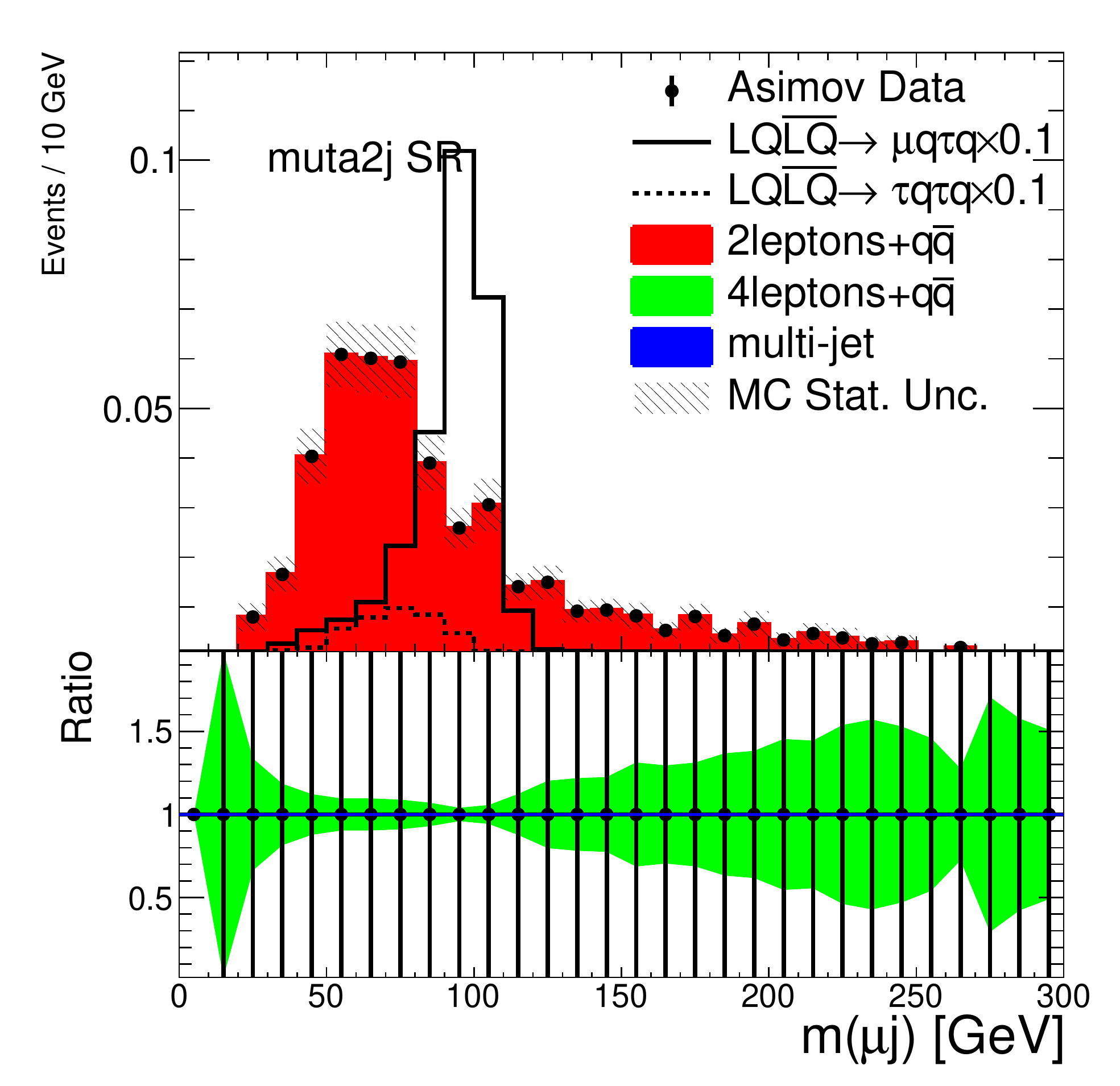}
    \includegraphics[width=0.35\textwidth]{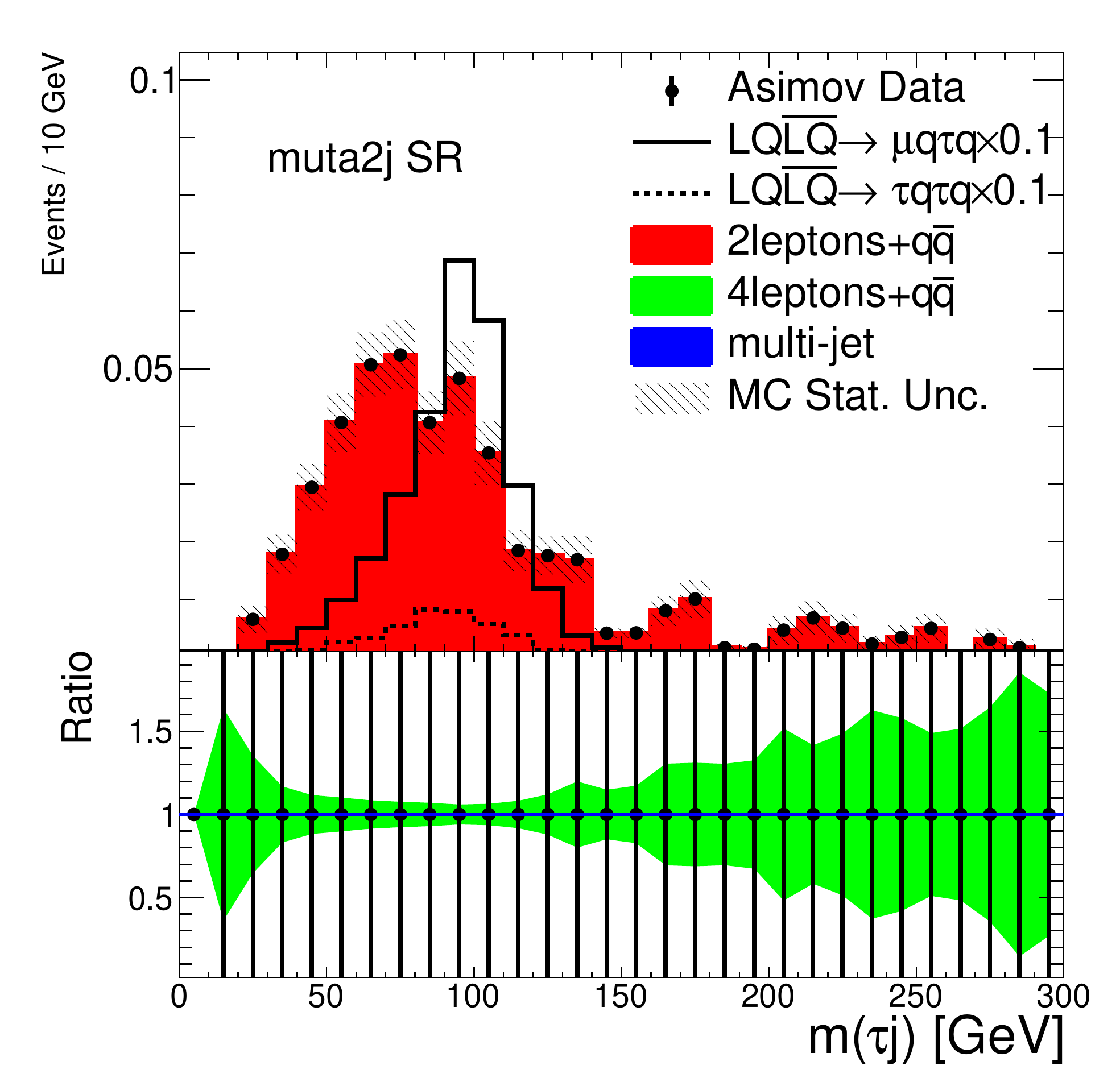}
    \caption{\label{fig:m_lq_muta2j}
        The distribution of the invariant mass of a lepton candidate and a jet candidate in the signal region ``muta1j''(Top) and ``muta2j''(Bottom).
        The left one is $m(\mu j)$ and the right one is $m(\tau j)$.
        The black points in the upper pads show the asimov data which is just the sum of all background events with uncertainties completely suppressed for better illustration. The ratio of the asimov data and the total background is shown in the lower pads, where the error bars on the black points represent the expected data uncertainty while the green bands represent the total MC statistical uncertainty. 
    }
\end{figure}
\begin{figure}[htbp]
    \centering
    \includegraphics[width=0.35\textwidth]{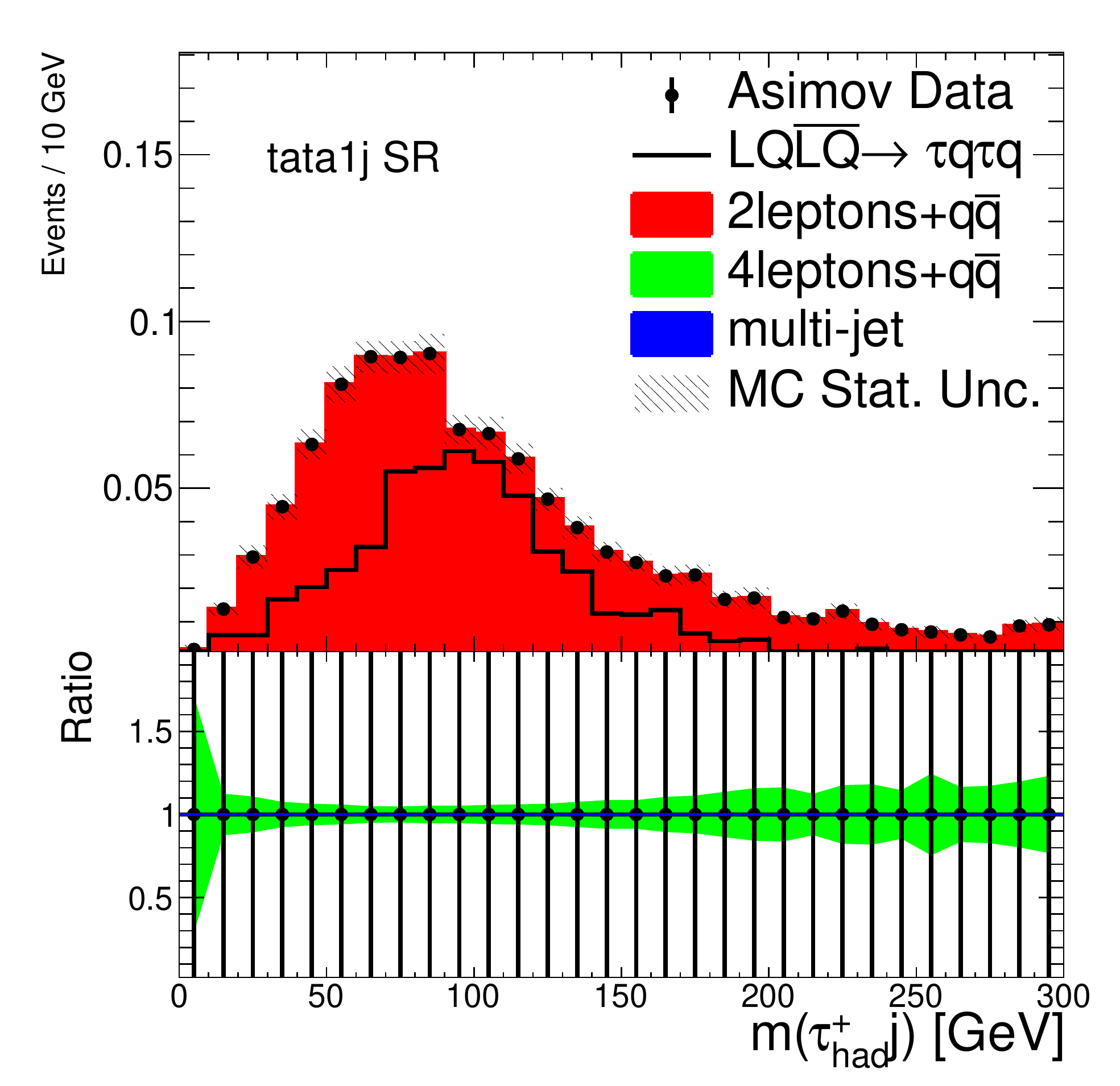}
    \includegraphics[width=0.35\textwidth]{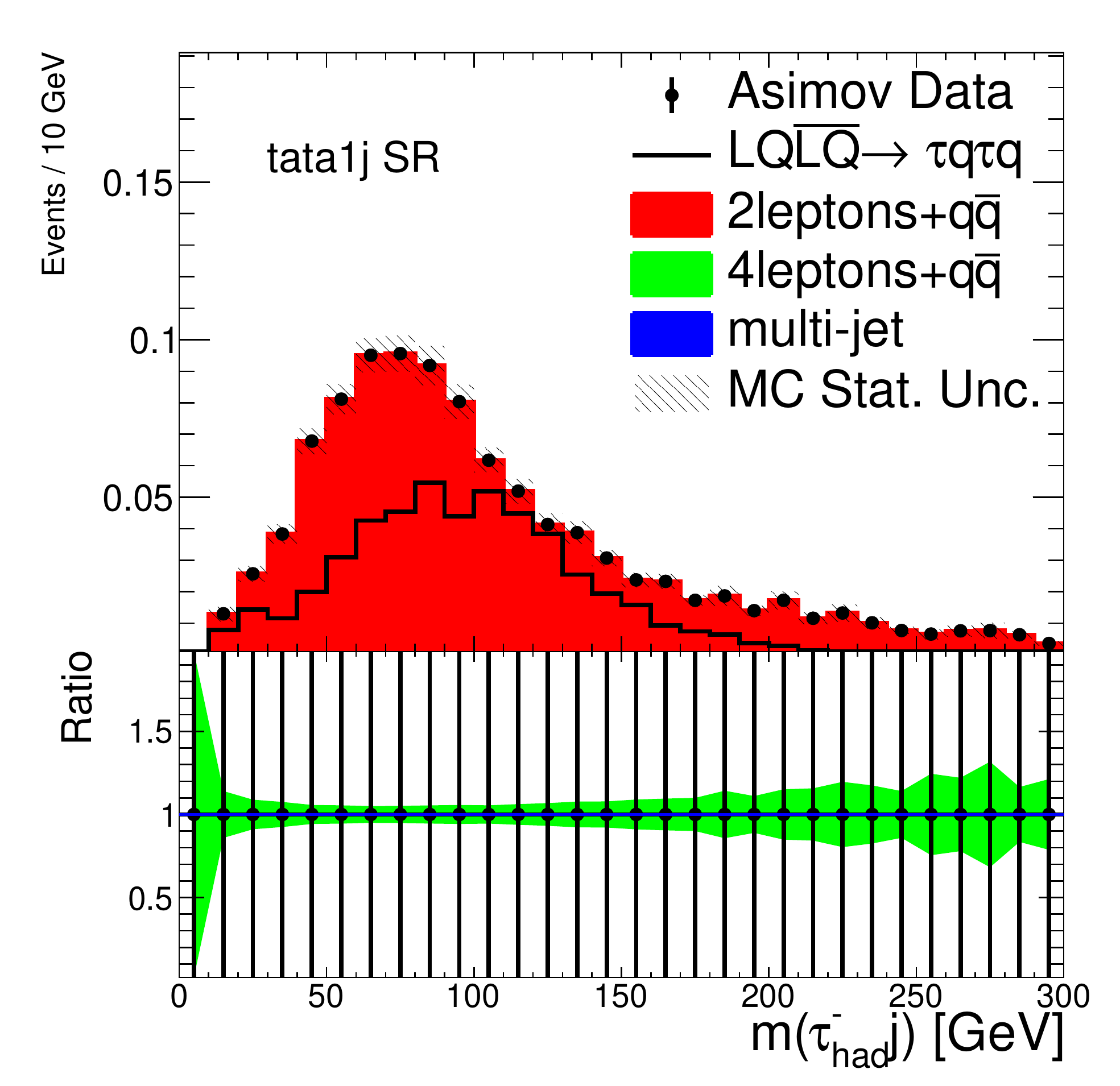}\\
    \includegraphics[width=0.35\textwidth]{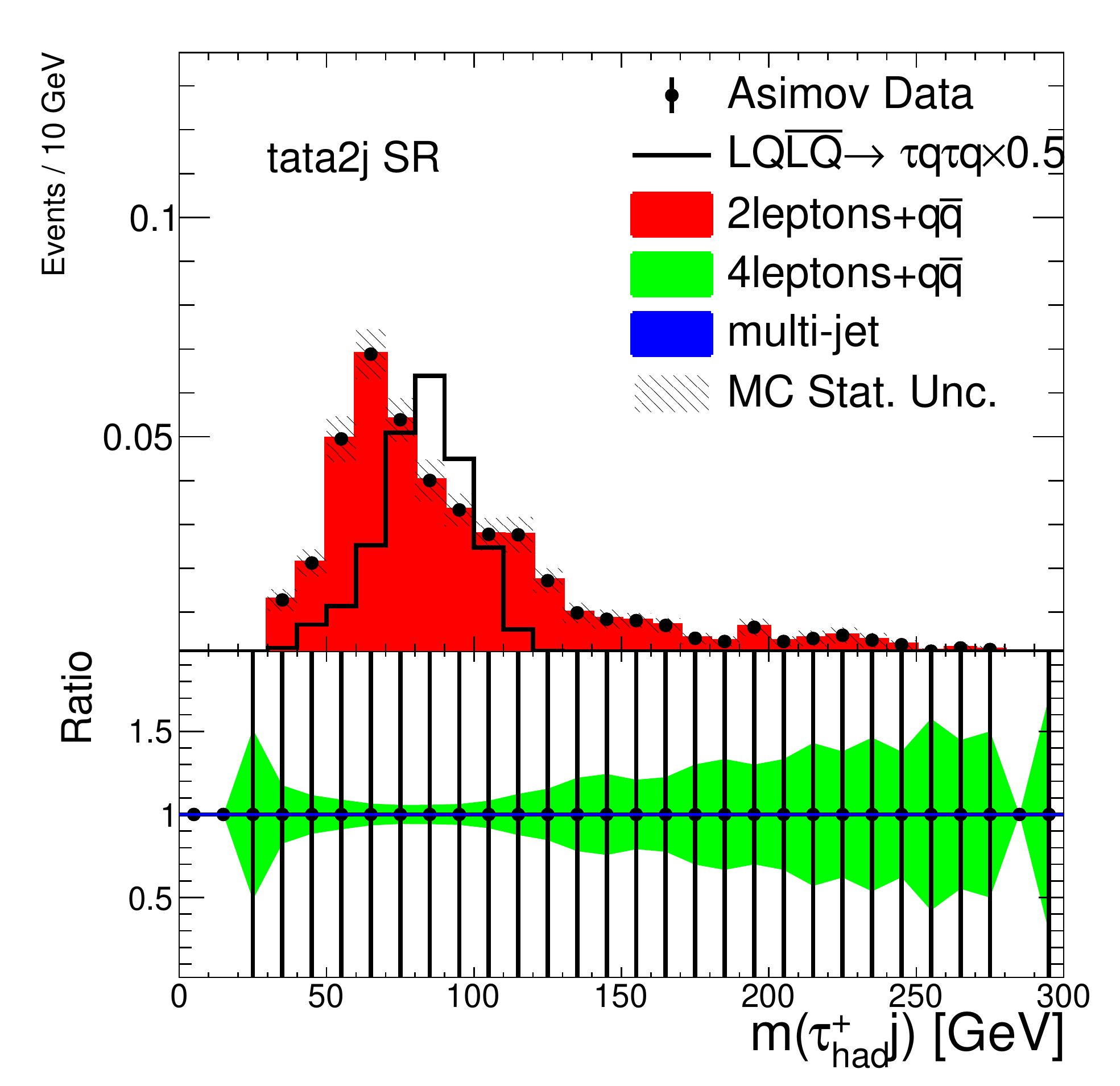}
    \includegraphics[width=0.35\textwidth]{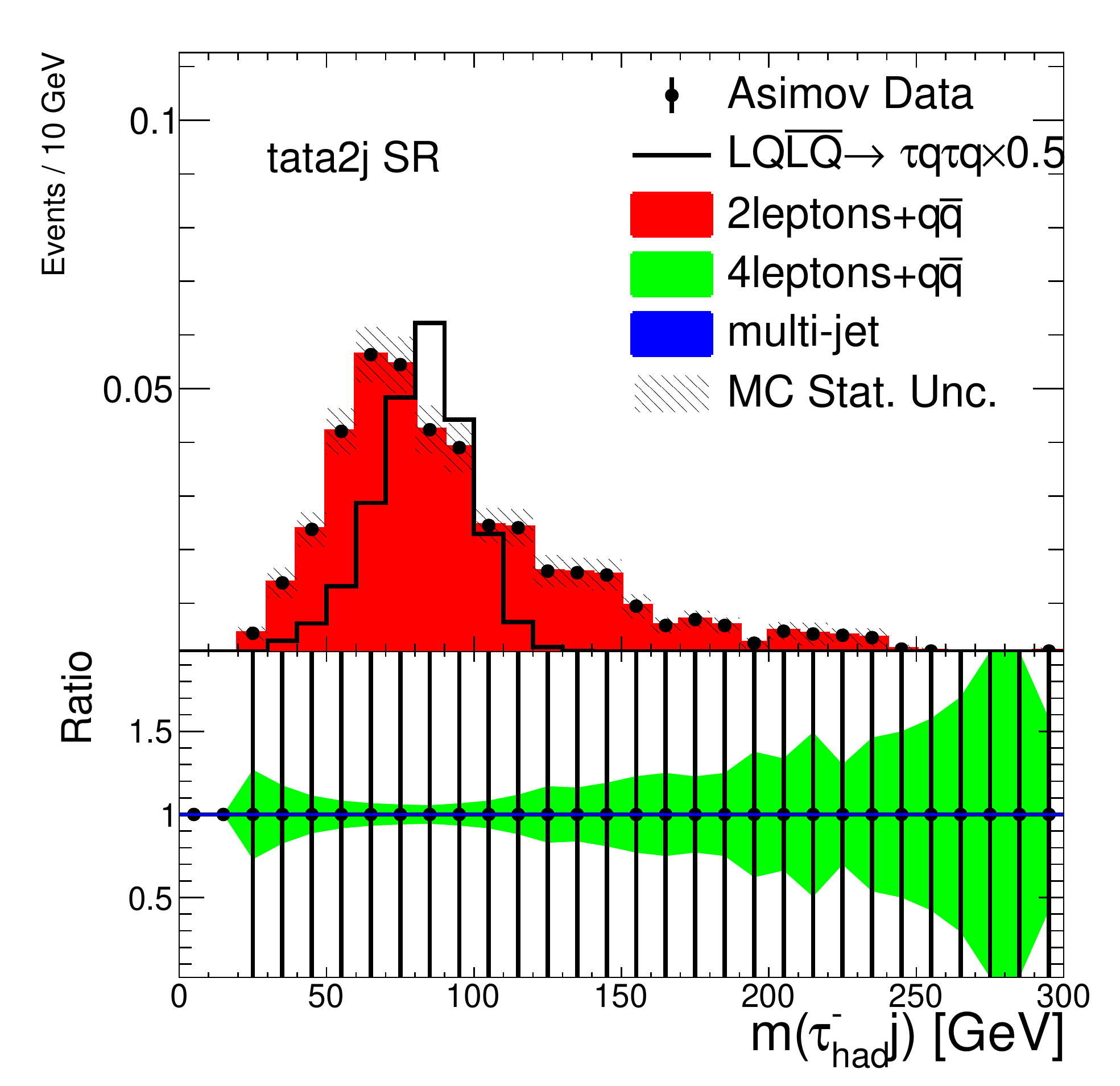}
    \caption{\label{fig:m_lq_tata2j}
        The distribution of the invariant mass of a lepton candidate and a jet candidate in the signal region ``tata1j''(Top) and ``tata2j''(Bottom).
        The left one is $m(\tauhad^+ j)$ and the right one is $m(\tauhad^- j)$.
        The black points in the upper pads show the asimov data which is just the sum of all background events with uncertainties completely suppressed for better illustration. The ratio of the asimov data and the total background is shown in the lower pads, where the error bars on the black points represent the expected data uncertainty while the green bands represent the total MC statistical uncertainty. 
    }
\end{figure}

Assuming a luminosity of 1~fb$^{-1}$, Fig.~\ref{fig:m_lq_ee2j}-\ref{fig:m_lq_tata2j} 
show the distribution of the invariant mass of a lepton and a jet candidates in all signal regions. We can see that the $S_3$ resonance is well reconstructed, which also confirms that the resonant production of a $S_3$ pair dominates. Other features in these distributions are also expected: 1) the mass resolution of $m(ej)$ and $m(\mu j)$ is better than that of $m(\tau j)$; 2) The resolution of $m(\tauhad j)$ is even worse because it is only a partial reconstruction of the $S_3$ resonance;
3) the background level in the emu2j signal region is quite low because lepton flavour is conserved in the SM. 

\begin{figure}[htbp]
    \centering
    \includegraphics[width=0.35\textwidth]{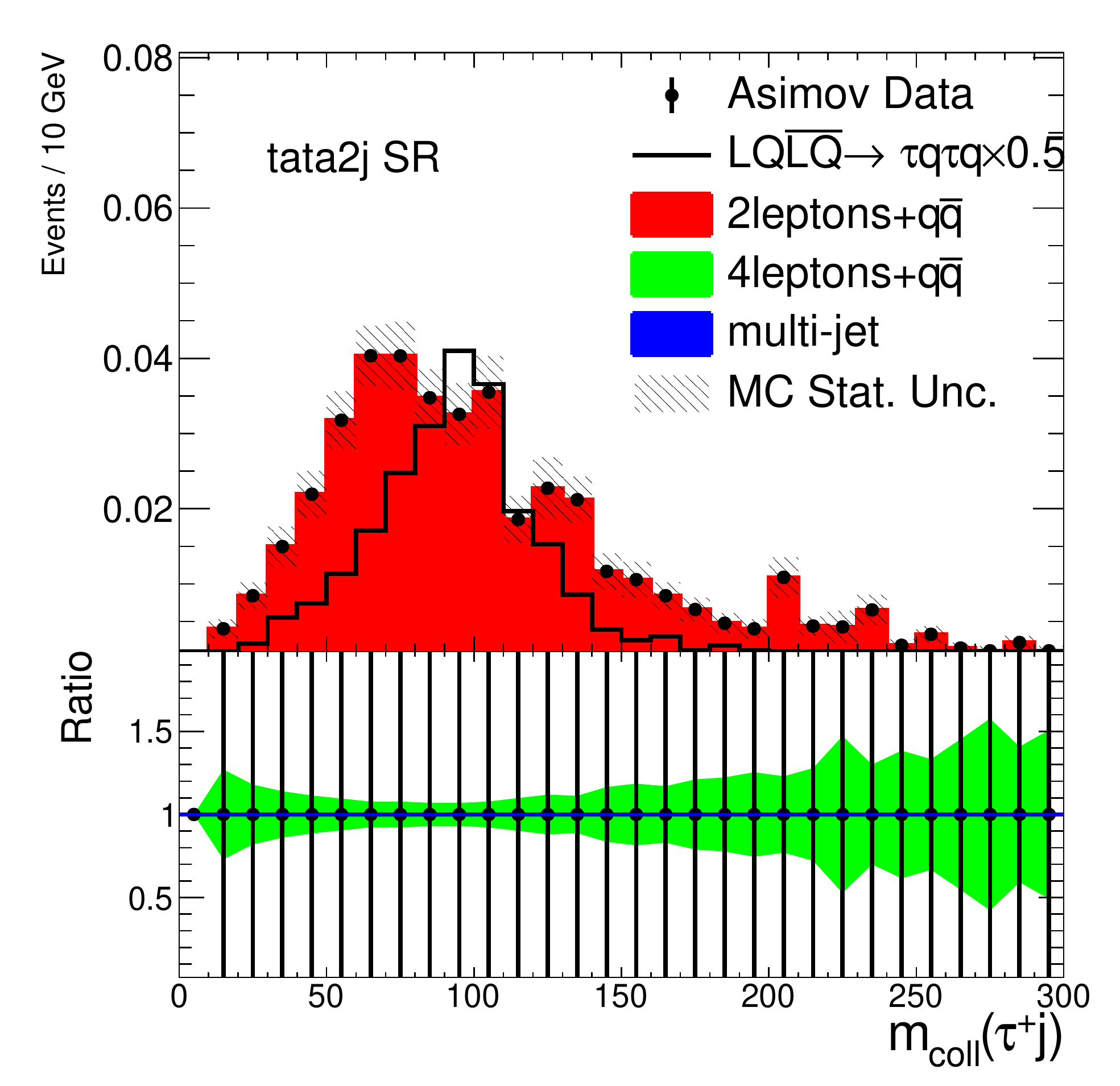}
    \includegraphics[width=0.35\textwidth]{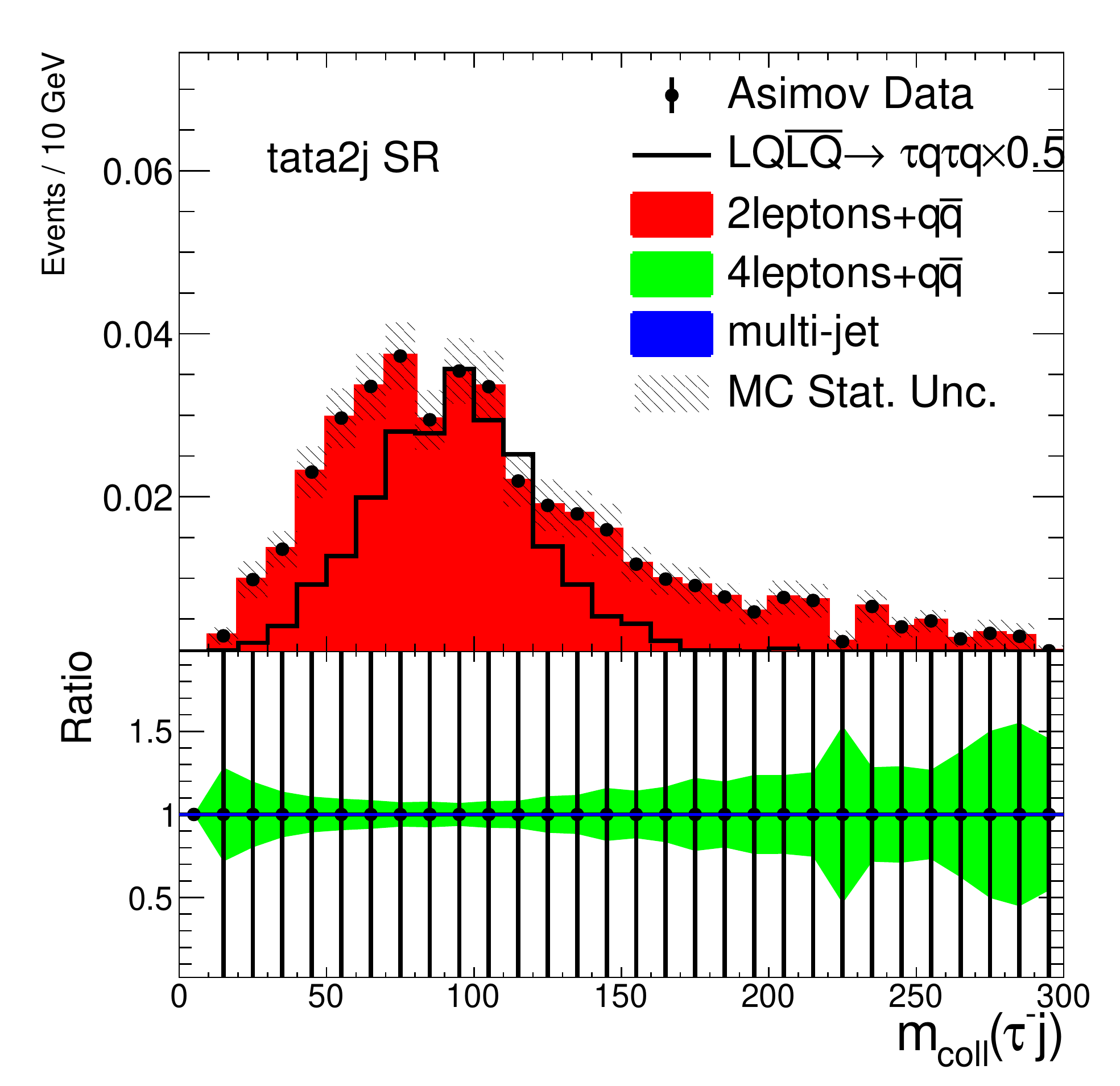}\\
    \includegraphics[width=0.35\textwidth]{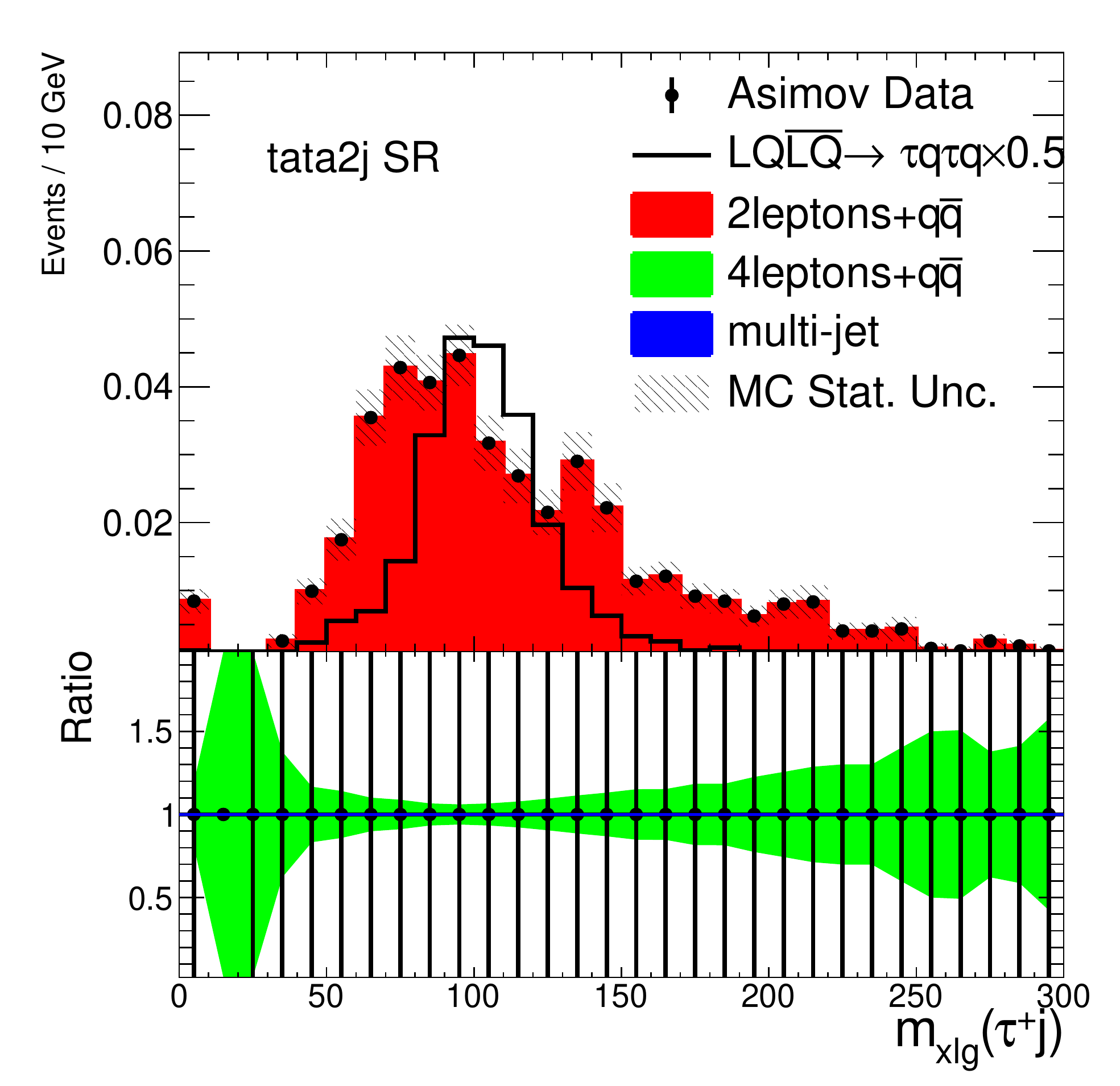}
    \includegraphics[width=0.35\textwidth]{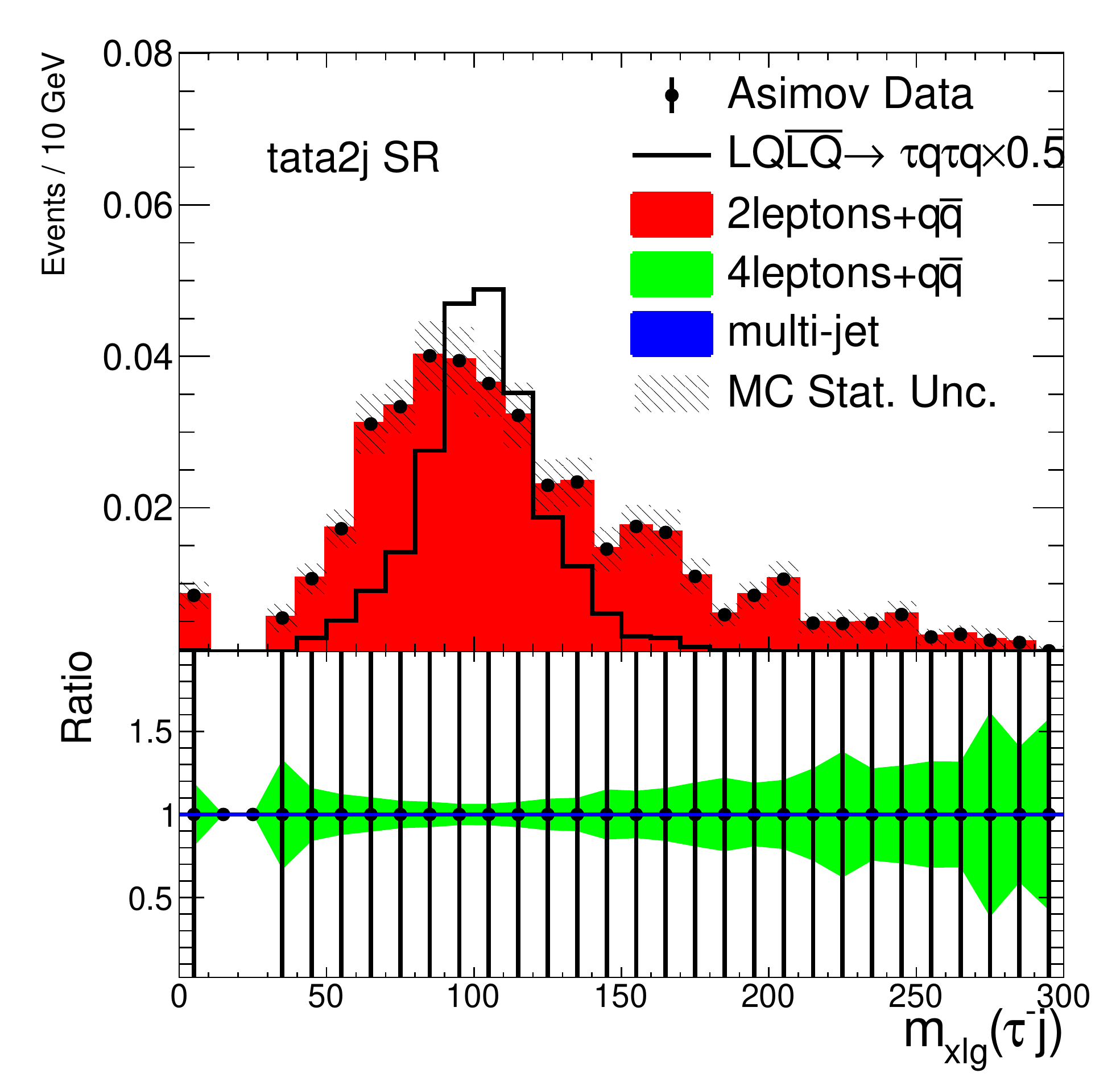}
    \caption{\label{fig:mcollmxlg_lq_tata2j}
        The distribution of the reconstructed mass of a lepton candidate and a jet candidate in the signal region ``tata2j'' based on the collinear approximation method (Top) and the method in Ref.~\cite{mxlg} (Bottom).
        The left one is $m(\tau^+ j)$ and the right one is $m(\tau^- j)$.
        The black points in the upper pads show the asimov data which is just the sum of all background events with uncertainties completely suppressed for better illustration. The ratio of the asimov data and the total background is shown in the lower pads, where the error bars on the black points represent the expected data uncertainty while the green bands represent the total MC statistical uncertainty. 
    }
\end{figure}

In order to improve the sensitivity to the mode $\LQ \to \tau q$, two mass reconstruction methods are tried in the tata2j signal region. 
One method is using the collinear approximation (technical details can be found in Ref.~\cite{mmc,mxlg}), the other method is from Ref.~\cite{mxlg} which reconstructs the mass by sampling the neutrino(s)' momentum distribution in the center-of-mass frame of the tau lepton. 
The two mass variables, $m_{\text{coll}}$ and $m_{\text{xlg}}$, are shown in Fig.~\ref{fig:mcollmxlg_lq_tata2j}. The latter mass $m_{\text{xlg}}$ has a better resolution and accurately reconstructs the $S_3$ mass. Thus it is used as the final observable for the tata2j signal region.



\section{Systematical uncertainties}\label{sec:unc}
The systematical uncertainties can be categorized into experimental uncertainties and theoretical uncertainties. Here is a summary of experimental uncertainties based on the ATLAS detector in Run~2 and main theoretical uncertainties.
\begin{itemize}
    \item Luminosity uncertainty: 4\%, which is quoted from Ref.~\cite{ATLAS_LbyL} following the method in Ref.~\cite{ATLAS_lumi_meas}.   
    \item Object triggering, reconstruction and identification efficiencies: the performance of final object triggering, reconstruction and identification can found in Ref.~\cite{ATLAS_electron_perf}, Ref.~\cite{ATLAS_muon_perf}, Ref.~\cite{ATLAS_tau_perf}, Ref.~\cite{ATLAS_met_perf}, and Ref.~\cite{ATLAS_jet_perf,ATLAS_jet_perf2} for electron, muon, hadronic tau, missing transverse energy and jet, respectively. 
        The efficiency uncertainties are usually expressed as a function of the transverse momentum. Table~\ref{tab:obj_rec_unc} shows the combined efficiency uncertainties (reconstruction and identification) for the transverse momentum range involved in this analysis. For jets, the dominant uncertainty is due to jet energy scale and energy resolution.
        In the transverse momentum range involved in this study, the systematical uncertainty is 5~\% on the energy scale and 6~\% on the energy resolution~\cite{ATLAS_jet_perf2}.
\begin{table}
    \centering
    \caption{\label{tab:obj_rec_unc}
        The combined efficiency uncertainty (reconstruction and identification) of the final objects. 
    }
        \begin{tabular}{l l}
            \hline
            Object & Combined efficiency uncertainty\\ 
            \hline
            Electron & about 7\%-3\% for $\pt=15-40$~GeV ~\cite{ATLAS_electron_perf}\\
            Muon &  about 1\% for $\pt=15-40$~GeV ~\cite{ATLAS_muon_perf}\\
            Hadronic tau& about 5\% for $\pt=20-60$~GeV ~\cite{ATLAS_tau_perf} \\
            \hline
        \end{tabular}
\end{table}
\item Photon flux: there are a number of theoretic uncertainties in the photon flux calculations. Some of them  need precise experimental measurements. One important uncertainty is about the probability of not having a hadronic interaction, $P_{0\text{had}}(b)$. As emphasized in the end of Sec.~\ref{sec:xs_UPC}, the different choices of $P_{0\text{had}}(b)$ show a photon flux uncertainty of 10-15\%~\cite{UPCreview1}. More discussions about these theoretic factors can be
    found in Ref.~\cite{UPCreview1,UPCreview2}. Photon flux uncertainty is directly related to the uncertainty of the signal production cross section. Here we present a data-driven estimation of this uncertainty. Because we have found a robust mass bound $m_{\LQ}>80$~GeV in Sec.~\ref{subsec:OPAL} and the irreducible background with $m(lj)<80$~GeV share the same final objects as the signal, thus the region with $m(lj)<80$~GeV can be used as a control region to constrain the photon flux uncertainty. 
\item Uncertainty of the background due to Pb-Pb hadronic interactions: the MC simulation of this background is unavailable and we have assumed it is negligible in the signal regions. 
    One of the key features of photon-photon interactions in UPCs is that the net transverse momentum of the final state is very small.
    This feature can be used to estimate the Pb-Pb hadronic background in the signal regions ``ee2j'', ``emu2j'', and ``mumu2j'' where no neutrinos are in the signal events. In other signal regions, we may use empirical functions to model it based on the fact that the background shape is expected to be smooth compared to the signal resonance structure.
\item Jet flavour: no attempt is made to identify the jet flavour. The jet reconstruction efficiency may differ for different flavours. This can be estimated by MC and the systematical uncertainty can be controlled by studying the process of $\gamma\gamma \to q\bar{q}$. A more detailed analysis by categorizing the events into signal regions with light jets (from $u,d,s$ quarks), $c$-tagged jets and $b$-tagged jets will surely improve the sensitivity. But the methodology is the same. 
\end{itemize}

\section{Statistical interpretation and sensitivity results}\label{sec:stats}
A simultaneous binned likelihood fit is performed to the distributions shown in Figs.~\ref{fig:m_lq_ee2j}-\ref{fig:mcollmxlg_lq_tata2j}. The likelihood function is 
\begin{equation}
    \mL(\kappa,B_{ex}, B_{\mu x}) = \Pi_{i=1}^{12} \Pi_{j=1}^{N_{\text{bins}}^i}P(n_{j}^i|b_j^i+\kappa \sum_{k=1}^{N_{\text{sig}}^i} s_{j,k}^i(B_{ex},B_{\mu x}))\:,
\end{equation}
where the index $i$ represents the $i$-th signal region; the index $j$ represent the $j$-th bin; the index $k$ represent the $k$-th signal mode; $N_{\text{bins}}^i$ represents the number of bin in the $i$-th signal region; $N_{\text{sig}}^i$ represent the number of signal modes in the $i$-th signal region; and $s_{j,k}^i(B_{ex},B_{\mu x})$ represents the yield contribution of the $k$-th signal mode to the $j$-th bin in the $i$-th signal region. 
The yield of each signal mode should be associated to the branching fractions appropriately. For example, we need to multiply the signal yield of $\LQ\overline{\LQ} \to eq\tau q$ by a factor of $2B_{ex}(1-B_{ex}-B_{\mu x})$. Furthermore, we should impose the constraint $0\leq B_{ex}+B_{\mu x}\leq 1$ as stated in Sec.~\ref{sec:method}.

\begin{figure}[htbp]
    \centering
    \includegraphics[width=0.45\textwidth]{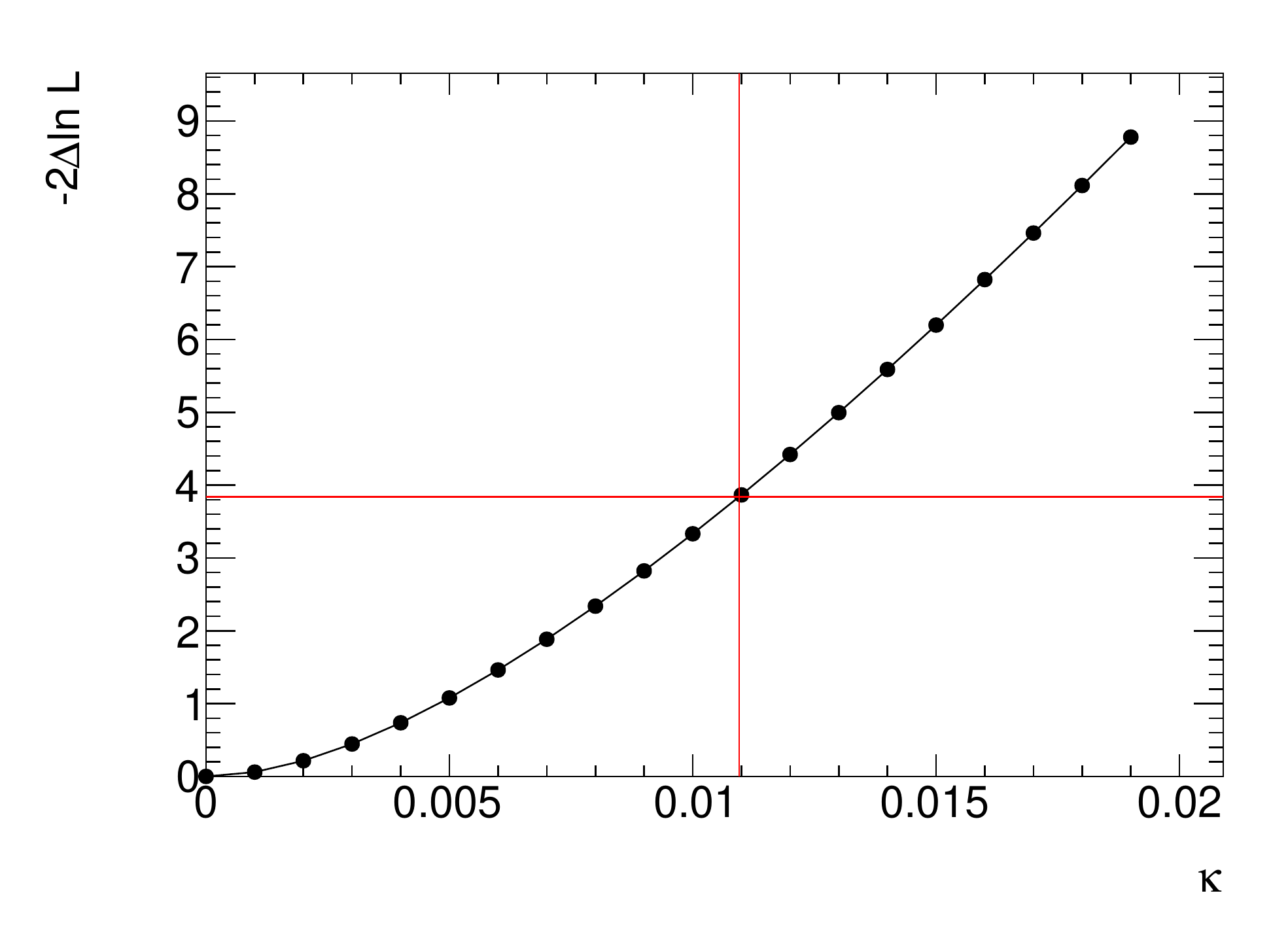}
    \includegraphics[width=0.45\textwidth]{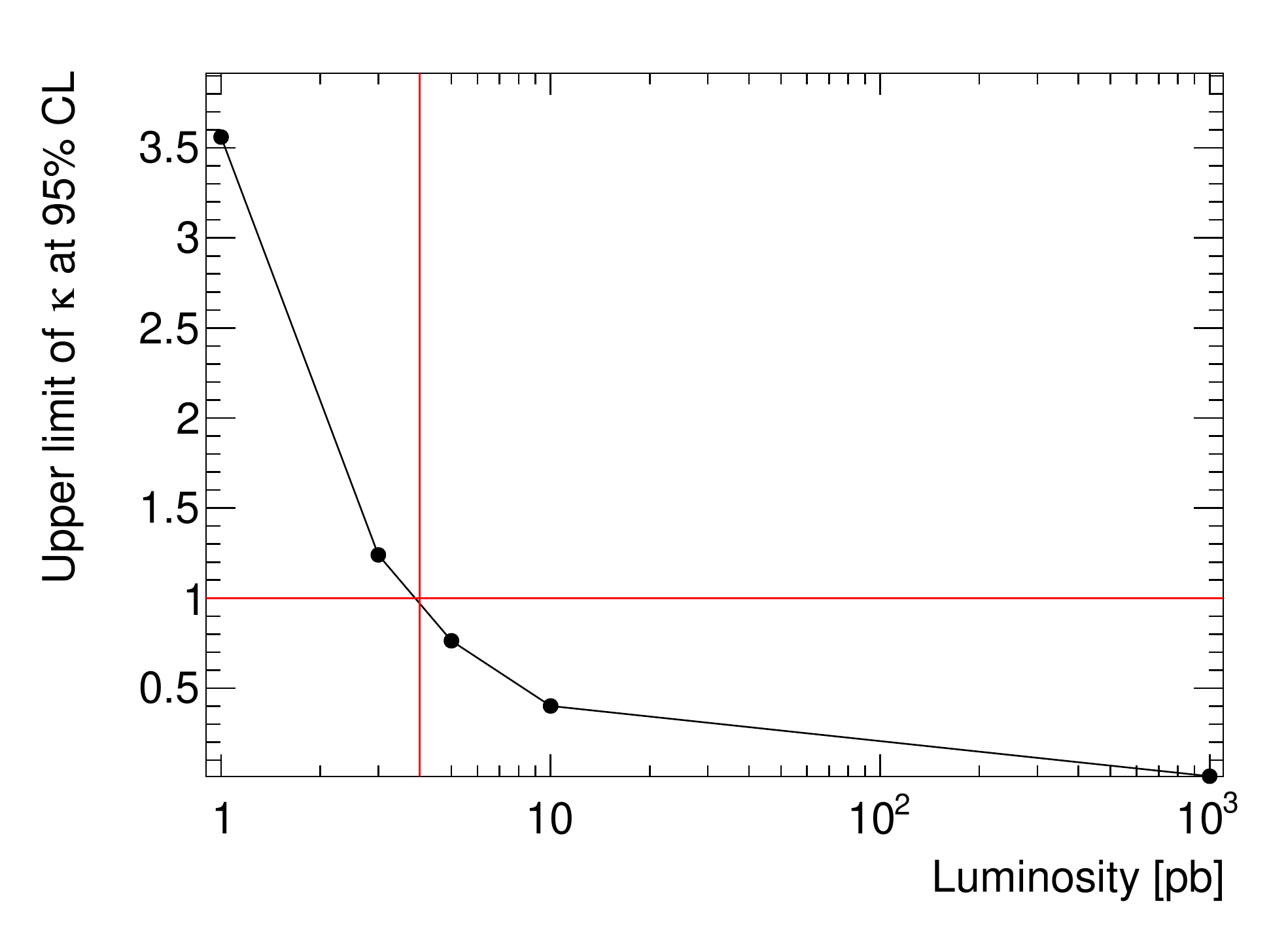}
    \caption{\label{fig:llscan_kappa}
    Left: the $-2\Delta\ln\mL$ as a function of $\kappa$ for a luminosity of 1~fb$^{-1}$. The red lines indicate the interval at 95~\% confidence level. Right: the upper limit of $\kappa$ at 95~\% confidence level as a function of the luminosity. The red lines indicate the luminosity corresponding to the upper limit of $\kappa$ being 1.
    }
\end{figure}

Allowing $B_{ex}$ and $B_{\mu x}$ to float, the left plot of Fig.~\ref{fig:llscan_kappa} shows the $-2\Delta\ln\mL(\kappa)\equiv -2\ln\frac{\mL(\kappa)}{\mL_{\text{max}}}$ as a function of $\kappa$ assuming the absence of new physics. We can see that $\kappa>1.1\times10^{-2}$ can be excluded at 95~\% confidence level for a dataset with the luminosity $L=1$~fb$^{-1}$. To exclude $\kappa>1$,
a dataset of $L\approx4.0$~pb$^{-1}$ is needed as shown in right plot of Fig.~\ref{fig:llscan_kappa}. The sensitivity results are summarized in Table~\ref{tab:results}.

\begin{table}
    \centering
    \caption{\label{tab:results}
    Expected sensitivity to search for $S_3$ with a mass of 100~GeV using 1~fb$^{-1}$ of Pb-Pb UPCs at $\sqrt{s}=5.02$~TeV.
    }
        \begin{tabular}{l l l }
            \hline
            $S_3$ mass &  Upper limit of $\kappa$ (95~\%C.L.) for 1~fb$^{-1}$ & Luminosity to exclude $\kappa=1$  \\
            \hline
            100~GeV & $1.1\times10^{-2}$ &  4.0~pb$^{-1}$ \\
        \hline
        \end{tabular}
\end{table}
\section{Can we exclude a 100~GeV LQ unambiguously?}\label{sec:ex_m100}
In this section, we allow a LQ to decay to all possible fermions and investigate the possibility of a 100~GeV LQ with an electric charge $\frac{4}{3}e$ considering all available experimental constraints. The results from the H1 experiment~\cite{H11,H12,H13}, CDF experiment~\cite{CDF0,CDF1,CDF2}, and ATLAS experiment~\cite{ATLAS2} are used. We do not include other experiments which did not look at this mass region although they may potentially provide stronger constraints.
Table~I summarizes the available constraints with the orignal interpretation with the corresponding assumption. We need to reinterpret them to remove the assumption. If an experiment does not explicitly identify the flavour of a jet, we will always include either all light down-type quarks (denoted by $q=d,s$) or all down-type quarks (denoted by $Q=d,s,b$) in the reinterpretation.  

\begin{table}[htbp]
    \centering
    \caption{\label{tab:constraints100}
    Reinterpretation of the experimental constraints on a 100~GeV LQ with an electric charge $\frac{4}{3}e$. In the table, $q$ denotes the light down-type quarks $d$ and $s$; $Q$ denotes all down-type quarks $d$, $s$ and $b$.
    }
        \begin{tabular}{l l l l}
            \hline
            Exp. & Source & Original constraint (assumption) & Re-interpretation \\
            \hline
            H1~\cite{H11} & Fig.~13 in ~\cite{H11} & $\lambda_{ed}\lesssim 0.01$ ($B_{ed}=1$) & $\lambda_{ed}^2B_{eq} < 0.01^2$\\ 
            H1~\cite{H12} & Fig.~3 in ~\cite{H12} & $\lambda_{ed/\mu d}\lesssim 0.001$ ($B_{ed/\mu d}=\frac{1}{2}$) & $\lambda_{ed}^2\frac{B_{\mu q}}{\frac{1}{2}} < (0.001)^2$\\
            H1~\cite{H13} & Fig.~3 in ~\cite{H13} & $\lambda_{ed}\lesssim 0.006$ ($B_{ed}=1$) & $\lambda_{ed}^2B_{eq}\lesssim (0.006)^2$\\ 
            CDF~\cite{CDF0} & Table~I in~\cite{CDF0} & $\sigma < 10$ pb ($B_{\tau Q}=1$) & $\sigma_{\text{th}}B_{\tau Q}^2<10$ pb, $\sigma_{\text{th}}=9.8$ pb \\
            CDF~\cite{CDF1} & Table~I in~\cite{CDF1} & $\sigma < 1.1$ pb ($B_{eq}=1$) & $\sigma_{\text{th}}B_{eq}^2<1.1$ pb, $\sigma_{\text{th}}=11.13$ pb \\
            CDF~\cite{CDF2} & Table~II in~\cite{CDF2} & $\sigma < 1.35$ pb ($B_{\mu q}=1$) & $\sigma_{\text{th}}B_{\mu q}^2<1.35$ pb, $\sigma_{\text{th}}=11.13$ pb \\
            ATLAS~\cite{ATLAS2} &  Fig.~4 in~\cite{ATLAS2} & Exclude 200-800~GeV (SG)  & see text \\ 
            \hline
        \end{tabular}
\end{table}

In the H1 experiment, a LQ is produced via the single production mode $e+ p \to \LQ +X$. For $\LQ \to lq$, the final yield is proportional to $\lambda_{ed}^2 B_{lq}$ (we ignore the possible interference with the SM process). The constraints are then re-expressed as
\begin{eqnarray}
    && \lambda_{ed}^2B_{eq}\lesssim (0.01)^2 \\ 
    && \lambda_{ed}^2 B_{\mu q}<2\times (0.001)^2\\
    && \lambda_{ed}^2B_{eq}\lesssim (0.006)^2 \:, 
\end{eqnarray}
which are from Ref.~\cite{H11,H12,H13}, respectively. 

In the CDF experiment, a pair of LQs is produced in the reaction of $p\bar{p}\to \LQ\overline{\LQ}+X$. The final yield is proportional to the product of the pair-production cross section and the branching fractions. Using the cross section from the theory, we can obtain the constraints on the branching fractions, namely, 
\begin{eqnarray}
    && B_{\tau Q} < 1.01 \\
    && B_{eq} < 0.31  \label{eq:CDF1}\\
    && B_{\mu q}< 0.35 \:, \label{eq:CDF2}
\end{eqnarray}
which are from Ref.~\cite{CDF0,CDF1,CDF2}, respectively.

Based on the constraints above, it is impossible to exclude a 100~GeV LQ definitely because all couplings are not covered. 
The ATLAS and CMS experiments have also performed comprehensive searches for various LQs via either single- or pair-production mechanism. However, we find only Ref.~\cite{ATLAS2} from ATLAS has a public dataset which can be used to probe the existence of a 100~GeV LQ with an electric charge $\frac{4}{3}e$.  The original motivation of Ref.~\cite{ATLAS2} is to search for a LQ decaying to $t\tau/b\nu$ or $b\tau/t\nu$. 
We will use their data with the final state $\taulep\tauhad$ plus one or more $b$-tagged jets, namely, Fig.~4 (middle left) from Ref.~\cite{ATLAS2} 
and Fig.~6b, Fig.~6c and Fig.~7b in its auxiliary material~\cite{ATLAS2aux}. The data with the $\tauhad\tauhad$ final state is not used due to the complicated trigger requirements which depend upon the real-time data-taking status. 

\begin{figure}[htbp]
    \centering
    \includegraphics[width=0.35\textwidth]{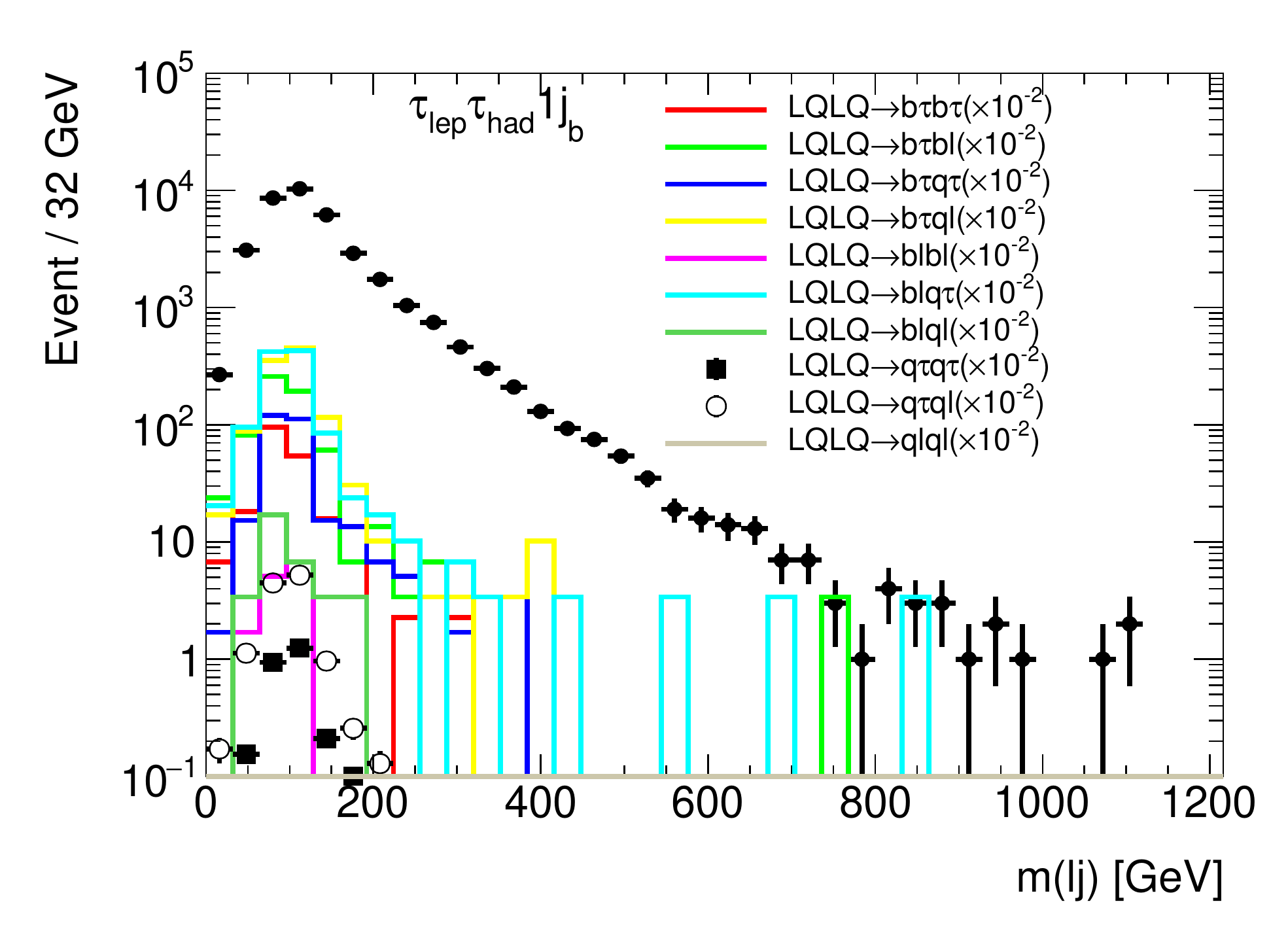}
    \includegraphics[width=0.35\textwidth]{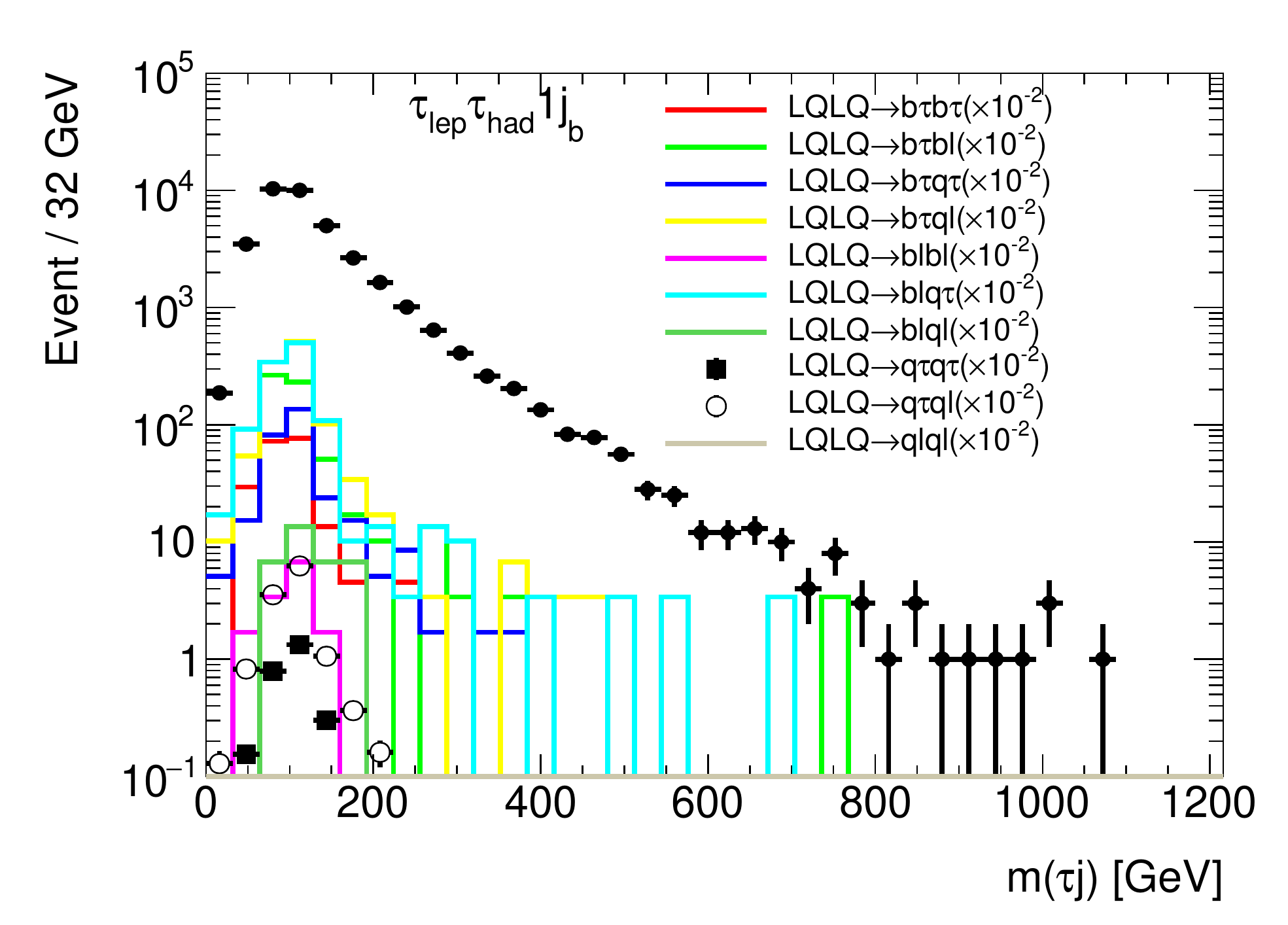}\\
    \includegraphics[width=0.35\textwidth]{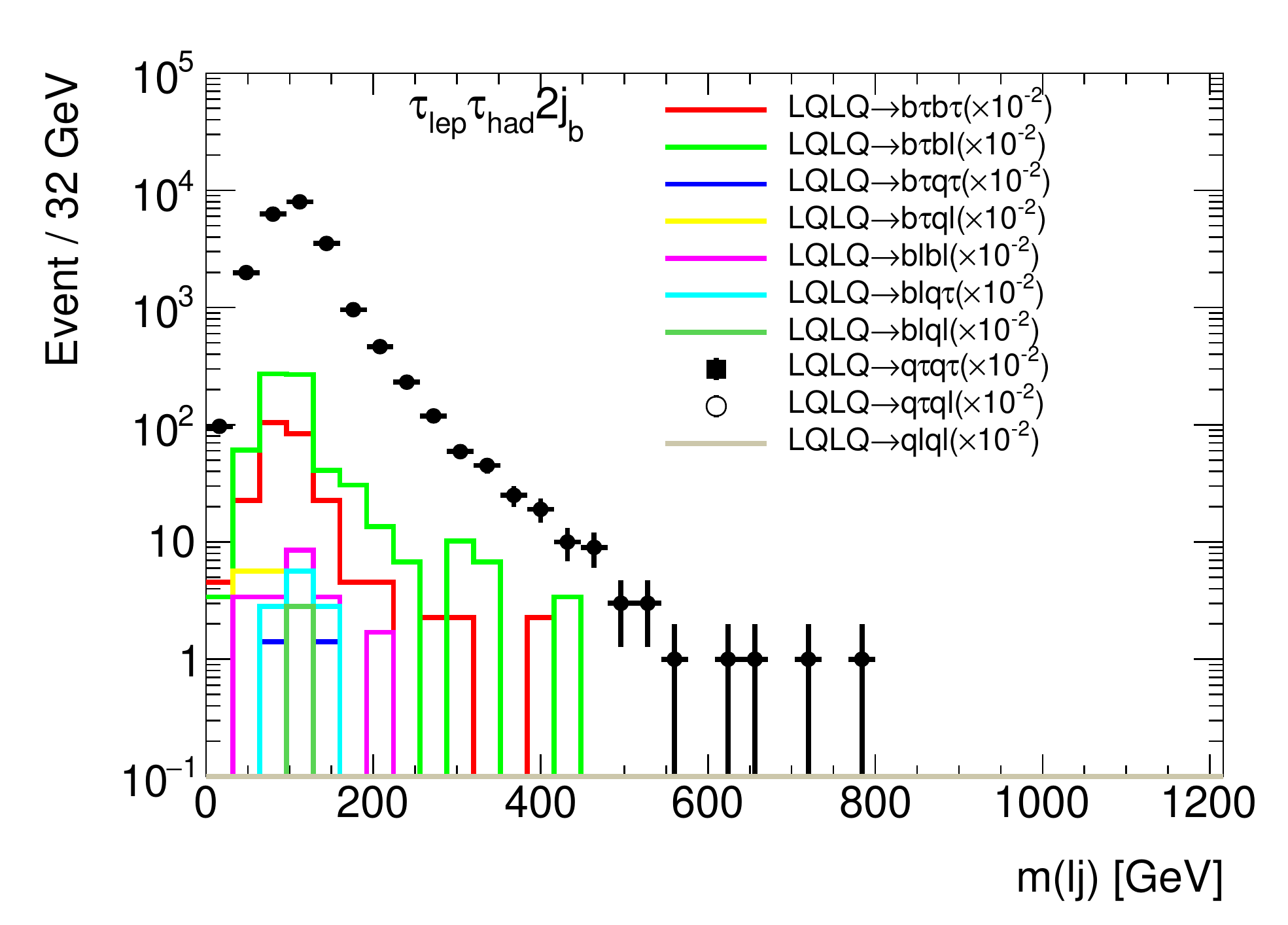}
    \includegraphics[width=0.35\textwidth]{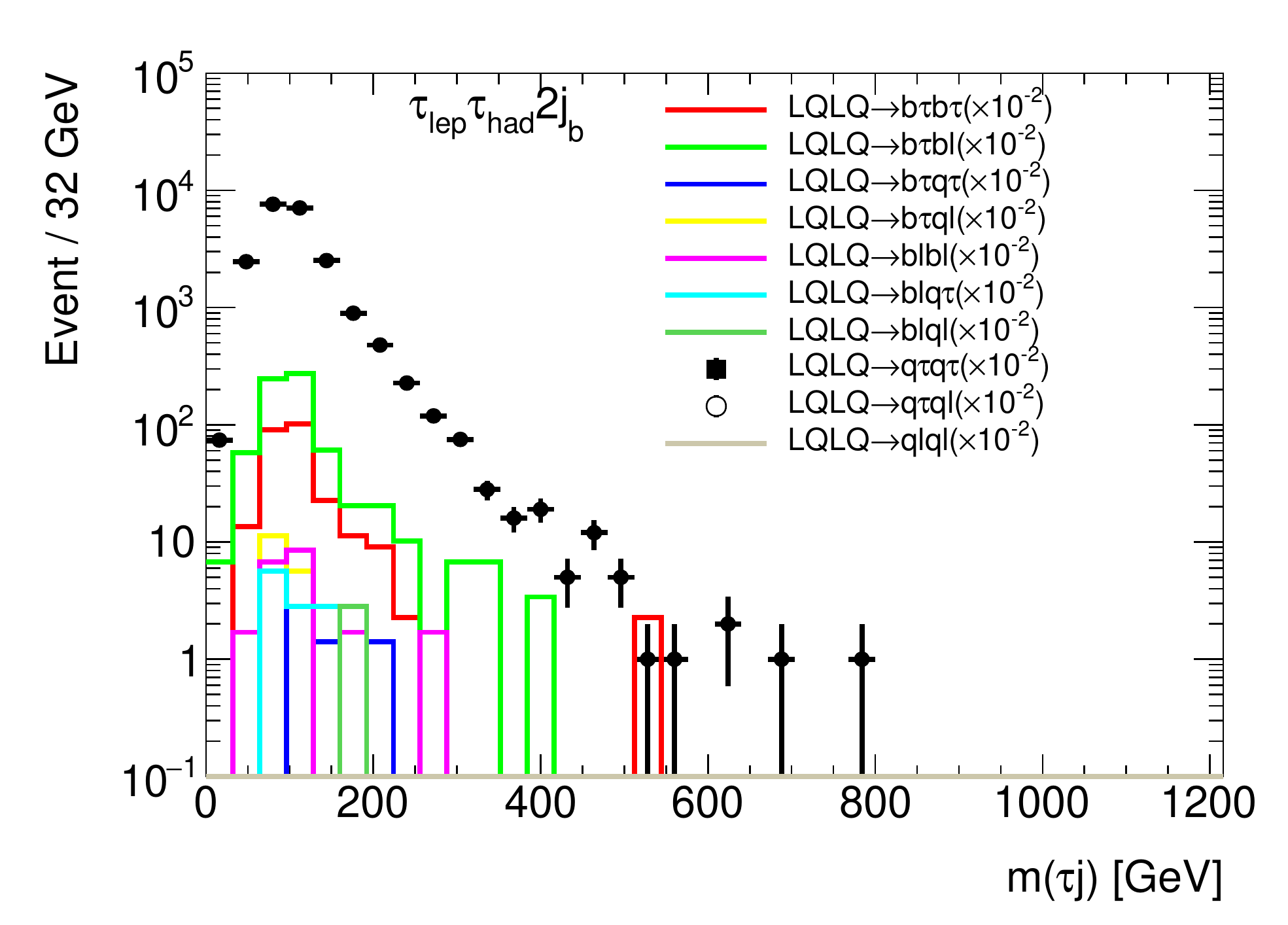}\\
    \caption{\label{fig:m_lq_ATLAS2}
        The distribution of the pair mass of a tau-lepton candidate and its mass-matched jet candidate in the signal region with one $b$-tagged jet (Top) and  two $b$-tagged jets (Bottom).
        The left one is $m(lj)$ and the right one is $m(\tau j)$.
        The black points represent the data from Ref.~\cite{ATLAS2,ATLAS2aux}. The open circles, black squares and histograms represent different signal topologies for a 100~GeV LQ. 
    }
\end{figure}

Figure~\ref{fig:m_lq_ATLAS2} shows the distribution of the pair mass of a tau-lepton candidate and its matched jet candidate in the signal region with one $b$-tagged jets (denoted by $\taulep\tauhad1j_b$) and signal region with two $b$-tagged jets
(denoted by $\taulep\tauhad2j_b$). The matching criterion is to minimize the mass difference of the LQ pair candidate (see Ref.~\cite{ATLAS2}).  

\begin{figure}[htbp]
    \centering
    \includegraphics[width=0.45\textwidth]{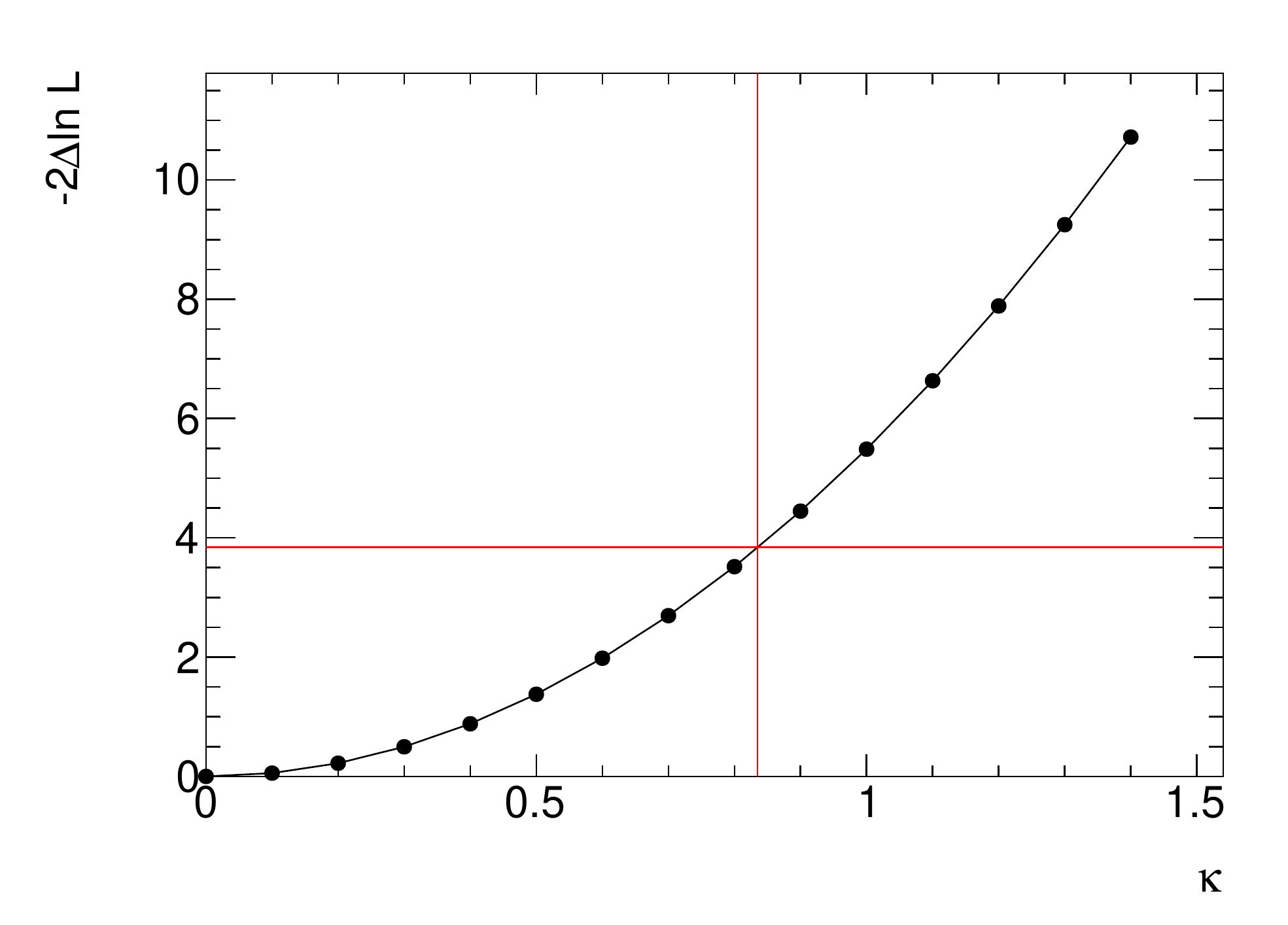}
    \caption{\label{fig:llscan_m100}
    The $-2\Delta\ln\mL$ as a function of $\kappa$ for a 100~GeV LQ. The red lines indicate the interval at 95~\% confidence level. 
    }
\end{figure}

We allow all possible decay modes, and do not differentiate $e$ and $\mu$ or $d$ and $s$. We thus have four decay modes, $\LQ \to b\tau,bl,q\tau,ql$ where $l=e,\mu$ and $q=d,s$ and 10 individual signal topologies from a LQ pair. They are also shown in Fig.~\ref{fig:m_lq_ATLAS2}. In preparing signal samples, we have tried our best to apply the same selection criteria as described in Ref.~\cite{ATLAS2}.  
Given that no significant deviation from the SM prediction is seen and following the methodology presented in this work, the upper limit at 95~\% C.L. of the pair-production cross section modifier $\kappa$ is determined to be 0.83 as shown in Fig.~\ref{fig:llscan_m100}, for which the CDF constraints in Eq.~\ref{eq:CDF1}-~\ref{eq:CDF2} have been considered. It means that a LQ with a mass 100~GeV is excluded at 95~\% C.L. under the general condition that a LQ can decay to all possible fermions. 

It is worthwhile to make two comments: 1) the sensitivity is dominated by the modes $q\tau q\tau$ and $q\tau ql$ (marked as squares and open circles in Fig.~\ref{fig:m_lq_ATLAS2}) for which we should design a signal region without any $b$-tagged jet in reality; 2) Ref.~\cite{ATLAS2} claims to exclude a LQ up to 1030~GeV assuming $B_{b\tau}=1$. It is likely that the exclusion region would shrink much dropping the assumption. Given the observation in this section, this work is better seen as a
proof of concept in order to search for LQs in a least model-dependent way.

\section{Summary}\label{sec:summary}
The possibility of low-mass leptoquarks cannot be excluded due to the assumptions made in previous experimental measurements.
In this work, we propose a least model-dependent method (assuming $\lambda\lesssim 0.1$) to search for them in the Pb-Pb UPCs thanks to the high photon flux and potentially low background contamination. It allows a LQ to couple to all possible lepton-quark pairs. 
Taking the scalar LQ $S_3$ as example, the feasibility is investigated using all $llqq$ final states based on the Pb-Pb UPCs at $\sqrt{s}=5.02$~TeV and the performance of the ATLAS detector in Run~2. 
The LQ with a mass of 100~GeV can be excluded at 95~\% confidence level using a dataset of 4.0~pb$^{-1}$.
This proposal uses the property that LQs carry electric charge and is complementary to the searches in $p$-$p$ collisions. The method also works for searching for high-mass LQs in $p$-$p$ collisions as long as the pair production mechanism dominates.

\acknowledgments
I am greatly indebted to David d'Enterria for support with MadGraph and the theoretical discussions about UPC. I also wish to express my appreciation to Prof. Shan Jin for helping to resume my research career. Especially I would like to thank Fang Dai for her encouragement and financial support.

\begin{appendix}
    \section{Signal purity comparison between in Pb-Pb collisions and in $p$-$p$ collisions}\label{app:purity}
    In this appendix, we compare the signal purity between in Pb-Pb collisions at 5.02~TeV and in $p$-$p$ collisions at 13~TeV using the $bb\tau\tau$ final state (an experimental search see Ref.~\cite{ATLAS_LQ3}). The final objects are reconstructed in the same way as described in Sec.~\ref{sec:selection}. The $b$-jet is reconstructed at a working point with an efficiency of 80~\%. The events with one charged lepton candidate ($e$ or $\mu$), one hadronic tau decay candidate and one $b$-jet candidate are selected. 
    The luminosity is 1~fb$^{-1}$ for Pb-Pb collisions and 36~fb$^{-1}$ for $p$-$p$ collisions. 
    
\begin{figure}[htbp]
    \centering
    \includegraphics[width=0.35\textwidth]{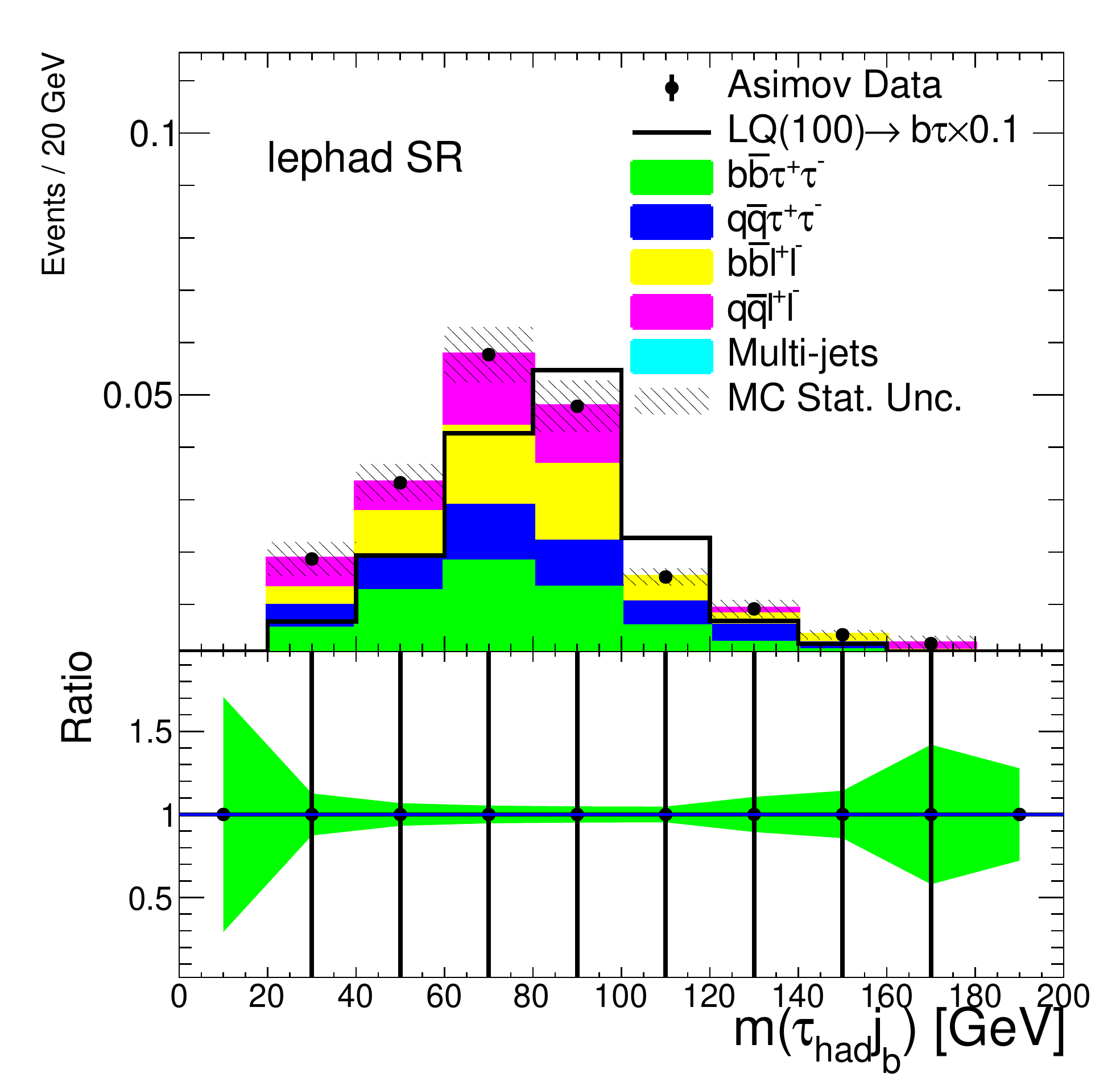}
    \includegraphics[width=0.35\textwidth]{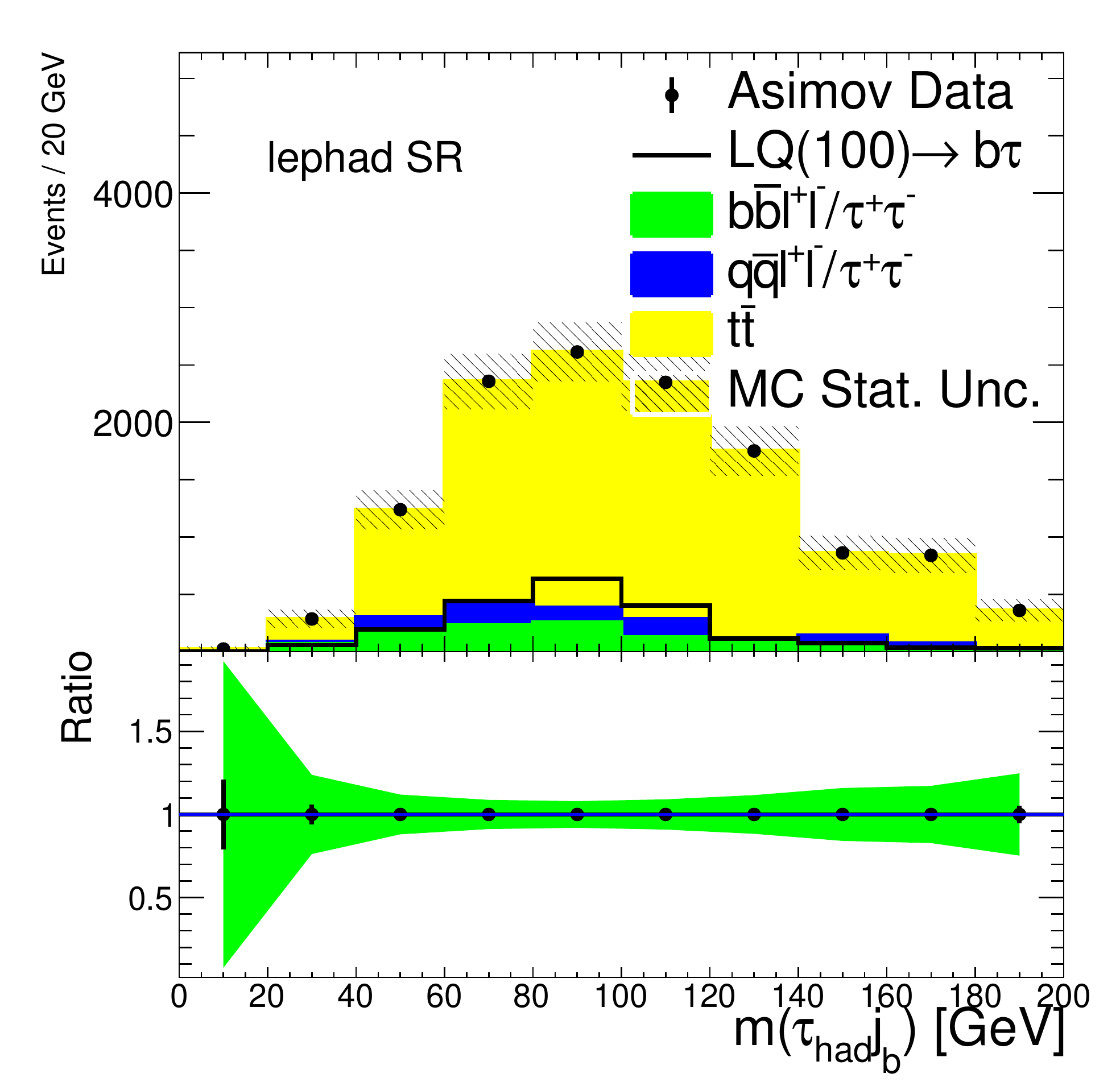}
    \caption{
        \label{fig:m_taujb}
        The distribution of the invariant mass of a hadronic tau candidate and a $b$-jet candidate in the Pb-Pb collisions (L) and the $p$-$p$ collisions (R). The signal histograms are scaled by a factor of $10^{-2}$ besides the scaling factor in the legends.
        The black points in the upper pads show the asimov data which is just the sum of all background events with uncertainties completely suppressed for better illustration. The ratio of the asimov data and the total background is shown in the lower pads, where the error bars on the black points represent the expected data uncertainty while the green bands represent the total MC statistical uncertainty. 
    }
\end{figure}

    Figure~\ref{fig:m_taujb} shows the distribution of the invariant mass of a hadronic tau candidate and a $b$-jet candidate. Both the production cross section in Pb-Pb collisions and in $p$-$p$ collisions are scaled by a factor $10^{-2}$ for better illustration (not including the scaling factor as shown in the legend in Fig.~\ref{fig:m_taujb}). We can see that the signal-to-background ratio in Pb-Pb collisions is around 50 times higher than that in $p$-$p$ collisions.
    The reason is that the $t\bar{t}$ background dominates in $p$-$p$ collisions while it is negligible in Pb-Pb collisions (the cross section is about 1~\% that of $bb\tau\tau$ production in the SM). This is one of the advantages of searching for low-mass LQs in Pb-Pb UPCs.

\end{appendix}

\end{document}